%% file: ms.tex
\documentclass[11pt,a4paper]{article}

\usepackage{jcappub}

\allowdisplaybreaks

\usepackage{ifthen}
\usepackage{graphicx}% Include figure files
\usepackage{dcolumn}% Align table columns on decimal point
\usepackage{bm}% bold math
\usepackage{color}
\usepackage{amsmath}
\usepackage{amssymb}
\usepackage{subfigure}
\usepackage{multirow}

\input{variables}

\begin{document}

\title{The Effect of High Column Density Systems on the Measurement of 
	the Lyman $\alpha$ Forest Correlation Function}

\input{authors}

\input{abstract}

\maketitle

\input{Introduction}

\input{Mocks}

\input{Results}

\input{Effect_Dlas}

\input{Metals}

\input{Conclusions}

\input{acknowledgements}

\bibliography{cosmo,cosmo_preprints}
\bibliographystyle{JHEP}

\appendix
\input{Peaks_bias}

\end{document}

%% file: variables.tex
\newcommand{\mpc}{\, {\rm Mpc}}

\newcommand{\hmpc}{\, h^{-1} \mpc}

\newcommand{\kms}{\, {\rm km\, s}^{-1}}

\newcommand{\lya}{Ly$\alpha$ }
\newcommand{\lyaf}{Ly$\alpha$ forest\ }

\newcommand{\vxperp}{\mathbf{x_\perp}}

\newcommand{\vx}{\mathbf{x}}

\newcommand{\vR}{\mathbf{r}}

\newcommand{\vxone}{\mathbf{x_1}}
\newcommand{\vxtwo}{\mathbf{x_2}}
\newcommand{\vRot}{\mathbf{r_{12}}}
\newcommand{\cm}{\, {\rm cm}}

%% file: authors.tex
\author[a,b,c]{Andreu Font-Ribera,}
\author[d,e]{Jordi Miralda-Escud\'{e}}
%\author[d,e]{Jordi Miralda-Escud\'{e},}
%\author[c,f]{Patrick McDonald}

\affiliation[a]{Institut de Ci\`{e}ncies de l'Espai (IEEC-CSIC), 
E. de Ci\`{e}ncies, Torre C5, Bellaterra, Catalonia, Spain.}
\affiliation[b]{Institute of Theoretical Physics, University of Z\"{u}rich, Winterthurstrasse 190, Zurich, Switzerland}
\affiliation[c]{Lawrence Berkeley National Laboratory, 1 Cyclotron Road, 94720 Berkeley, California, U.S.A.}
\affiliation[d]{Instituci\'o Catalana de Recerca i Estudis Avan\c cats,
Catalonia}
\affiliation[e]{Institut de Ci\`{e}ncies del Cosmos (IEEC/UB),
Barcelona, Catalonia}
%\affiliation[f]{Brookhaven National Laboratory, Long Island, US}

\emailAdd{font@ieec.uab.es}

%% file: abstract.tex
\abstract{ 
 We present a study of the effect of High Column Density (HCD) systems on the 
 \lya forest correlation function on large scales. We study the effect both
 numerically, by inserting HCD systems on mock spectra for a specific model,
 and analytically, in the context of two-point correlations and linear theory.
 We show that the presence of HCDs substantially contributes to the noise of
 the correlation function measurement, and systematically alters the measured
 redshift-space correlation function of the \lya forest, increasing the value
 of the density bias factor and decreasing the redshift distortion parameter
 $\beta_\alpha$ of the \lya forest.
 We provide simple formulae for corrections on these derived
 parameters, as a function of the mean effective optical depth and bias factor
 of the host halos of the HCDs, and discuss the conditions under which these
 expressions should be valid. In practice, precise corrections to the measured
 parameters of the \lya forest correlation for the HCD effects are more complex
 than the simple analytical approximations we present, owing to non-linear
 effects of the damped wings of the HCD systems and the presence of three-point
 terms. However, we conclude that an accurate correction for these HCD effects
 can be obtained numerically and calibrated with observations of the
 HCD-\lya cross-correlation. We also discuss an analogous formalism to treat
 and correct for the contaminating effect of metal lines overlapping the \lya
 forest spectra.
}

\keywords{cosmology: large-scale structure ---  
          cosmology: spectroscopic surveys ---
 galaxies: intergalactic medium --- galaxies: absorption systems }

%% file: Introduction.tex
\section{Introduction}

  Observations of the correlation function of the \lya forest in
redshift space from multiple spectra is emerging as a powerful tool to
explore the large-scale structure of the universe at high redshift. This
development has been led by the BOSS survey, part of the SDSS-III
collaboration \cite{2011AJ....142...72E}, which is obtaining optical
spectra of 160,000 quasars at $z>2.1$ for the principal purpose of studying
the \lya forest absorption and measuring its power spectrum. The
redshift space power spectrum of the fluctuations in the fraction of
transmitted flux, $F$, has a complex form on small scales that is
affected by non-linear gravitational evolution, thermal broadening,
the non-linear relation between $F$ and the optical depth, and complex
physical processes such as galactic winds. But on large scales, the
power spectrum should be simply related to the mass power spectrum in
the linear regime, $P_L$, through two biasing parameters
\cite{2003ApJ...585...34M}:
\begin{equation}
 P_\alpha(k,\mu_k) = b_\alpha^2 (1+\beta_\alpha \mu_k^2)^2 \, P_L(k) ~,
\end{equation}
where $k$ and $\mu_k$ are the modulus and angle cosine relative to the
line of sight of the wave vector in redshift space, $b_\alpha$ is the
bias factor relating the amplitude of fluctuations in $F$ to the
relative amplitude of density fluctuations, and $\beta_\alpha$ is the
redshift distortion parameter. This form of the linear power spectrum in
redshift space is the same as that for discrete tracers of the density
field \cite{1987MNRAS.227....1K}, 
except that $\beta_\alpha$ depends also on the bias parameter for the
peculiar velocity gradient, $b_\eta$. Recently, the first measurement
of $b_\alpha$ and $\beta_\alpha$ for the \lya forest was reported by 
\cite{2011JCAP...09..001S} from the first year of BOSS data, and more accurate
measurements are expected in the near future.

  The values of $b_\alpha$ and $\beta_\alpha$ as a function of redshift
can be predicted in principle from numerical simulations of the \lya
forest \cite{2003ApJ...585...34M,2009JCAP...10..019S}, and they depend on the
detailed small-scale physical processes in the intergalactic medium.
Comparison of the predicted values with the observed ones will therefore
test these physical processes.
However, in practice the observed absorption spectra are affected not 
only by the low-density gas producing the \lya forest, but also by 
higher density systems that give rise to absorption lines of high column 
density, observed as Lyman limit systems (hereafter LLS, with column
densities $N_{HI} > 10^{17.2}\cm^{-2}$) and damped \lya systems (
hereafter DLA, with $N_{HI}>10^{20.3}\cm^{-2}$). 
These systems, as well as the lower column density \lya forest,
produce also metal absorption lines, some of which appear in the 
region of the \lya absorption and add to the contamination of the
measurement of the \lya power spectrum.

The presence of high column density systems (hereafter referred to as
HCDs, meaning both LLS and DLAs) has a similar effect on the \lya power
spectrum as 
the well-known ``fingers of God'' in galaxy redshift surveys: on small, 
non-linear scales, galaxies accumulate in high-density clusters with an 
internal velocity dispersion, appearing in redshift 
space as highly elongated structures along the line of sight. This
induces contours of the correlation function that are 
also elongated along the line of sight on small scales, precisely 
the opposite to the squashing effect on the correlation function contours 
induced by the Kaiser linear term in the power spectrum that
is prevalent on large scales. In the 
case of absorption spectra, the damped wings of the HCDs may similarly
spread the correlation function along the line of sight.
%and metal lines with wavelengths overlapping the \lya forest region can
%also have a similar effect.
However, contrary to the ``fingers of God''
in galaxy surveys, the effects of damped wings extend out to all large
scales in the \lya forest, owing to their power-law absorption profiles.
Metal lines can also cause an elongation of contours when they overlap
the \lya forest and introduce bumps in the correlation function near
the line of sight around the separation that corresponds to the
wavelength difference between the metal and the \lya lines. In addition
to this effect, there is also the purely linear fact that if the HCDs
have a different redshift distortion factor than the \lya forest, the
correlation of the combined transmission will display an averaged
redshift distortion factor of the absorber populations that are
contributing to the total absorption.

While most previous work studied the effect of HCDs on the power spectrum 
along the line-of-sight \cite{2005MNRAS.360.1471M,2004MNRAS.349L..33V},
this paper focuses on the impact of HCDs on the linear bias factors of the
\lya forest. Their effect on the measured power spectrum is determined
by the fact that HCDs are correlated with the underlying mass
distribution and therefore with the \lya forest intergalactic
absorption. Their presence also adds additional noise to any power
spectrum measurements. The impact of metal-line absorbers is also
important and was briefly discussed in \cite{2011JCAP...09..001S}. 
Here we present a
description of the expected effect, and we describe a method to correct
it in the \lya correlation measurements.

  In Section \ref{sec:mockdla} we present a method to introduce HCD systems in
\lya mock spectra. The effect of HCD in the measurement of the \lyaf 
correlation inferred from mocks is presented in Section \ref{sec:results}. 
An analytical description of this effect is described in Section
\ref{sec:effectHCD}. Finally, the impact of metal lines is 
discussed in Section \ref{sec:metals}.

A standard flat $\Lambda CDM$ cosmology is used in this paper with the
following parameters:
$h=0.72$ , $\Omega_m=0.281$, $\sigma_8=0.85$, $n_s=0.963$, $\Omega_b=0.0462$.

%% file: Mocks.tex
\section{Model for the High Column Density systems}
\label{sec:mockdla}

  The impact of HCDs on the correlation function of \lya absorption
depends on their column density and Doppler parameter distribution, and
on the way they are distributed in space relative to the underlying \lya
forest. In this section, we describe the method we use to introduce
these systems in mock \lya absorption spectra.
We first briefly summarize the method to generate
the \lya forest mock spectra \cite{2012JCAP...01..001F}. Then we describe our
model distributions for HCDs, and the way they are inserted in the mock
spectra with a correlation with the \lya absorption field.

\subsection{\lya mock spectra}

  The reader is referred to \cite{2012JCAP...01..001F} for a full account of the
method we use to generate mock \lya spectra with any specified
three-dimensional flux power spectrum and flux probability distribution
function. Here we highlight the features that are most important for
this paper. The method consists of two steps:
\begin{itemize}
 \item A Gaussian random field $\delta_g(x)$ is generated for the
set of specified lines of sight.
 \item The field is transformed to a new variable $F(\delta_g)$
  constrained to the range $0 < F < 1$, determined by the condition
  of matching the model probability distribution of $F$.
  The power spectrum for the Gaussian variable $\delta_g$ is chosen
so that the final flux power spectrum of $F$ is the desired one.
\end{itemize}

  In general, as described in \cite{2012JCAP...01..001F}, a third step
can be applied where one interpolates the
value of $F$ between lines of sight generated at different redshifts to
simulate the effect of redshift evolution and the fact that the lines
of sight are not parallel. This third step is not included in this
paper. The lines of sight are generated as parallel lines at a fixed
redshift of $z=2.3$, with a value of the mean transmission fixed at
$F_\alpha = 0.791$, in order to study the effect of the HCD systems
without introducing additional complexities into the mocks.

We use here the same flux power spectrum $P_\alpha(k,\mu_k)$ 
from \cite{2003ApJ...585...34M}, 
\begin{equation}
 P_\alpha(k,\mu_k) = b_\alpha^2 \, (1+\beta_\alpha \mu_k^2)^2 \,
    P_L(k) \, D(k,\mu_k) ~,
 \label{eq:Pk}
\end{equation}
where $P_L(k)$ is the linear matter power spectrum, 
$\mu_k$ is the cosine of the angle of the Fourier vector from the 
line of sight, and $D(k,\mu_k)$ is a small scale non-linear term
that was fitted to the results of numerical simulations
in \cite{2003ApJ...585...34M}. 
We use the central values for the model parameters from the first row
of Table 1 in \cite{2003ApJ...585...34M}, after applying a small
correction to the bias parameter for the difference in redshift
(the biases in \cite{2003ApJ...585...34M} are computed at $z=2.25$)
by assuming that the power spectrum amplitude evolves as $(1+z)^{3.8}$
\cite{2011JCAP...09..001S}, which implies, neglecting the influence
of the cosmological constant on the growth factor at this redshift,
that the bias $b_{\alpha}$ evolves as $(1+z)^{2.9}$.
The resulting bias parameters are $b_\alpha=-0.1375$ and $\beta_\alpha=1.58$.
%Note that the negative value of
%$b_\alpha$ is a consecuence of the definition of 
%$\delta_\alpha = \frac{F_\alpha}{\bar F_\alpha}-1$, 
%that is positive in underdense regions of the universe.

\subsection{Column density distribution and Doppler parameters}

  A large part of the contamination by HCD systems on the \lya forest
correlation arises from the damped wings, which depend exclusively on
the column density. It is therefore most important to use a model that
reproduces the observed column density distribution of these systems.
The large number of quasars observed in the Sloan Digital Sky Survey
has in recent years allowed good determinations of this distribution
\cite{2005ApJ...635..123P,2009A&A...505.1087N}.

\begin{figure}[h!]
 \begin{center}
  \subfigure{\includegraphics[scale=0.4, angle=-90]{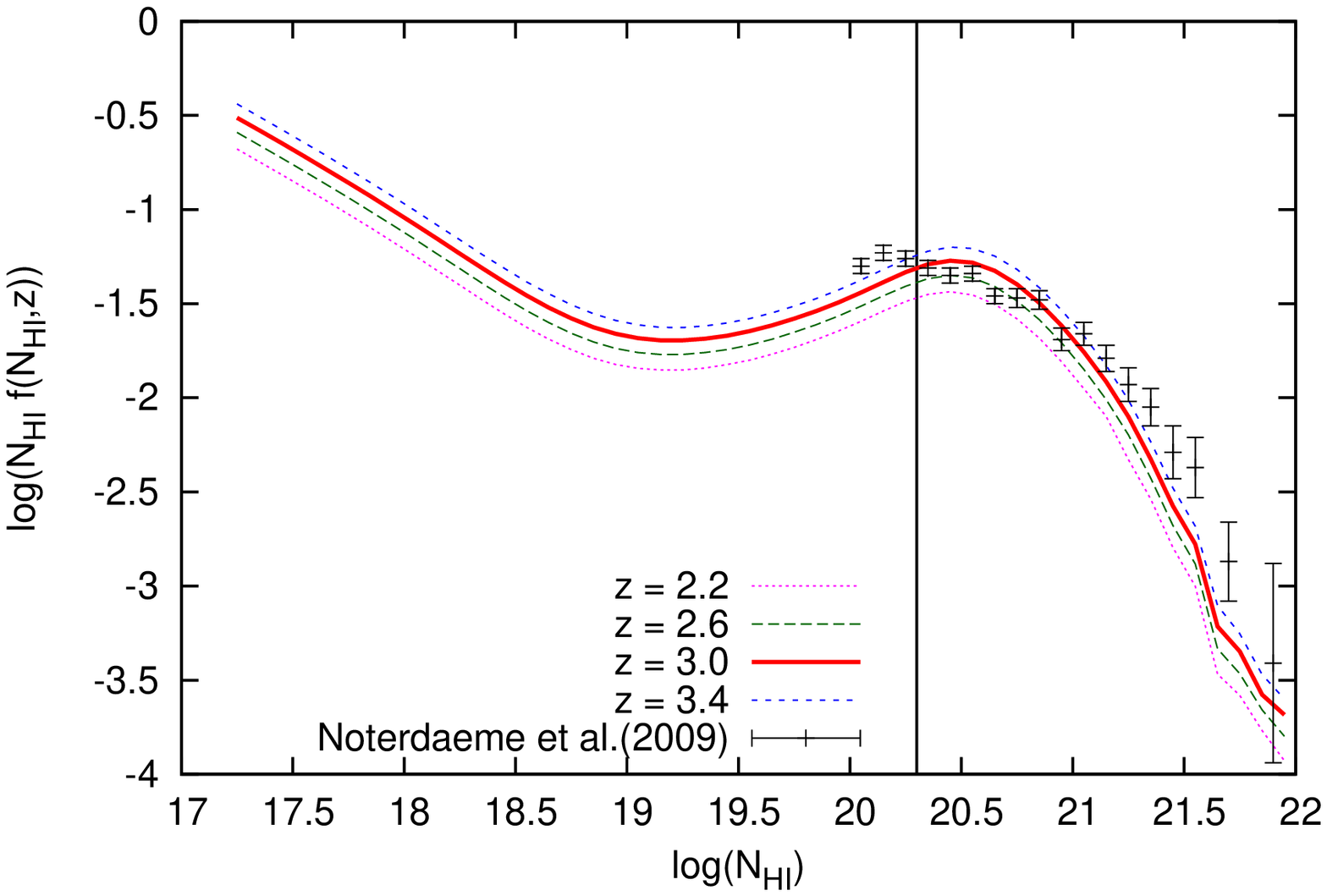}}
  \subfigure{\includegraphics[scale=0.4, angle=-90]{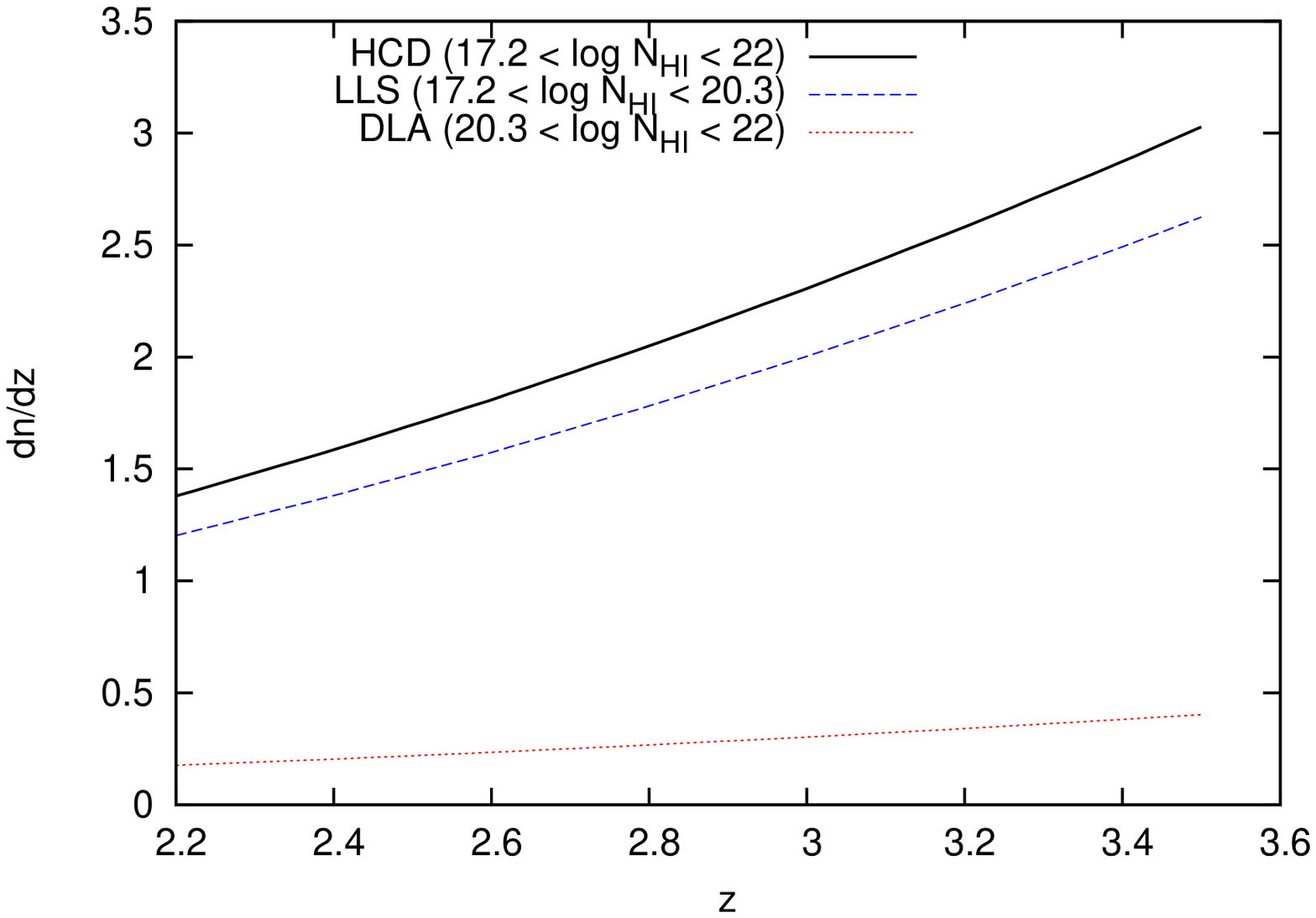}}
 \end{center}
 \caption{{\it Left}: Number of HCD systems per unit of column density
 $N_{HI}$ and unit of "absorption distance" as defined in the text. The lines
 show the values for our model, at $z=2.2$, $z=2.6$, $z=3$ and $z=3.4$.
 The vertical line indicates the standard separation between DLA and
 LLS. Points with errorbars show the observational determination in
 \cite{2009A&A...505.1087N}, with a central redshift of $z \approx 3$.
 {\it Right}: Number of systems per unit of redshift in the indicated
 column density ranges as a function of redshift.}
 \label{fig:column}
\end{figure}

Here we use the neutral hydrogen column density distribution used in
\cite{2005MNRAS.360.1471M}, which is based on an analytical expression
derived in \cite{2002ApJ...568L..71Z} that assumes an intrinsic
power-law distribution of the total hydrogen column density and takes
into account the self-shielding effects on the neutral column density,
and is calibrated to match the observations of DLAs in
\cite{2005ApJ...635..123P}.
In figure \ref{fig:column}, the column density distribution in this
model is shown at different redshifts ({\it left panel}), together with
the observations of \cite{2009A&A...505.1087N}. 
In this figure we plot the frequency distribution per unit of column density
and "absorption distance" X, defined as
\begin{equation}
 dX \equiv \frac{H_0}{H(z)}(1+z)^2 ~  dz ~ ,
\end{equation}
since this is the function presented in the SDSS analysis.
Self-shielding causes the flattening of the
distribution in the column density range of $10^{18}$ to
$10^{20}\cm^{-2}$. In the right panel we plot the number of systems as
a function of redshift integrated over various column density ranges.

\begin{figure}[h!]
 \begin{center}
  \includegraphics[scale=0.4, angle=-90]{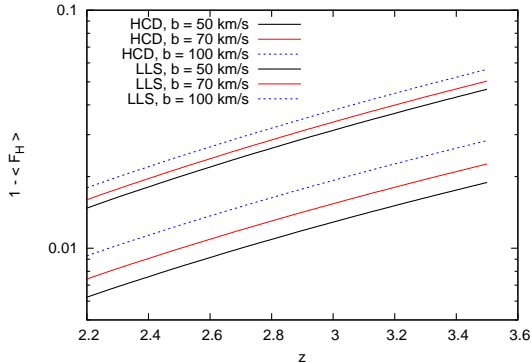}
 \end{center}
 \caption{Mean absorption caused by HCD systems for several values of $b_D$.
Lower lines include systems with $N_{HI} < 10^{20.3}\cm^{-2}$ only; upper
lines include all systems.}
 \label{fig:FH}
\end{figure}

%{\it From Jordi:} No se com escriure la resta d'aquesta subseccio,
%exactament per que mostres la figura 2? Quins valors de $b_D$ fas servir
%a la resta del paper? El parametre $b_D$ afecta nomes la part central
%del perfil, i per tant afectara la correlacio en funcio de l'efecte
%d'aquesta part central. Per que es important saber l'absorcio promig?
%Es aixo el que s'ha d'explicar.

  As discussed in section \ref{sec:effectHCD}, the perturbation caused by HCD 
systems on the \lya correlation increases with their contribution to the 
mean effective optical depth, $\bar\tau_{eH} = - \log (\bar F_H)$, which
depends on the velocity dispersions in 
addition to the column densities of the systems. Here we calculate its value as 
a function of redshift for different values of the Doppler parameter $b_D$, 
and the column density distribution used in this study.
Defining $W(N_{HI},b_D)$ as the rest-frame equivalent width according to
the standard curve of growth of an absorber with a Gaussian distribution
of velocities, the mean absorbed fraction by HCDs (assuming their positions
are uncorrelated) is
\begin{equation}
  \bar\tau_{eH}(z) = \int d N_{HI} ~ \frac{d^2n(z,N_{HI})}{dN_{HI} dz} ~
        \frac{W(N_{HI}, b_D)}{\lambda_\alpha} ~ (1+z) ~ .
\end{equation}

  This effective optical depth is plotted in Figure \ref{fig:FH} as a function
of redshift, separately for all the systems (HCD) and including only
systems with $N_{HI} < 10^{20.3}\cm^{-2}$ (designated here as LLS),
for three different values of $b_D$. The contribution from systems that
are not included in the definition of DLAs to $\bar\tau_{eH}$ is about
half of the total, and increases with $b_D$. At the redshift of our
mocks, the total effective optical depth from HCDs is close to 2\%.

  In the mocks of this paper we use a value of $b_D=70 \kms$, a
representative value for DLAs (see \cite{1997ApJ...487...73P}). We note that
the value of $b_D$ actually has a large dispersion and its mean
depends on the column density (being smaller for lower column density
systems). We shall not examine the possible dependence of the effects
on the correlation function we study on the distribution of $b_D$,
but we note that most of these effects arise from the damped wings,
which are unaffected by the velocity dispersion. Better observational
constraints on $\bar\tau_{eH}$ in the future will help quantify
the effect of HCDs more accurately.

\subsection{Clustering of HCDs}

  While the effect of HCDs on increasing the noise in the measurement
of the \lya correlation can be adequately estimated by simply placing
the absorption systems randomly in the mock spectra, the systematic
effect on the correlation is induced only by their clustering.
On large scales, this systematic effect on the total measured
correlation should be governed by the bias factors of the HCDs.

  Here we use a simple method to insert these systems in the mock
spectra, by placing them only in a fraction $\nu$ of pixels where the
optical depth is above a certain threshold, $\tau > \tau_c$.
For a fixed distribution function of $F=\exp(-\tau)$, the value of
$\nu$ determines the critical optical depth $\tau_c$
or, equivalently, a critical transmitted flux fraction $F_c$:
\begin{equation}
  \nu = \int_{\tau_c}^\infty d\tau ~ p_\tau(\tau) 
      = \int_{0}^{F_c} d F ~ p_F(F) 
\end{equation}
The dependence of the probability distribution function $p_\tau$ on
redshift implies that the threshold $\tau_c$ for hosting a HCD depends
also on redshift. In this paper we do not include redshift evolution to
avoid complications, and we analyze the effect of HCDs at the single
redshift $z=2.3$, although our method generally incorporates redshift
evolution when detailed mocks of the BOSS data are required. Here we
generally use $\nu=0.01$ (which yields $\tau_c \approx 12$ %=11.63 
at $z=2.3$), except for one model where we increase the HCD bias by
changing to $\nu=0.002$.

%\footnote{In \cite{XiPush} the systems were introduced at constant 
%abundance in comoving separation (calibrated at $z=2.6$), causing an 
%overabundance of systems at low redshift and an underabundance at high 
%redshift.}

\begin{figure}[h]
 \begin{center}
\subfigure{
	\includegraphics[scale=0.45, angle=-90]{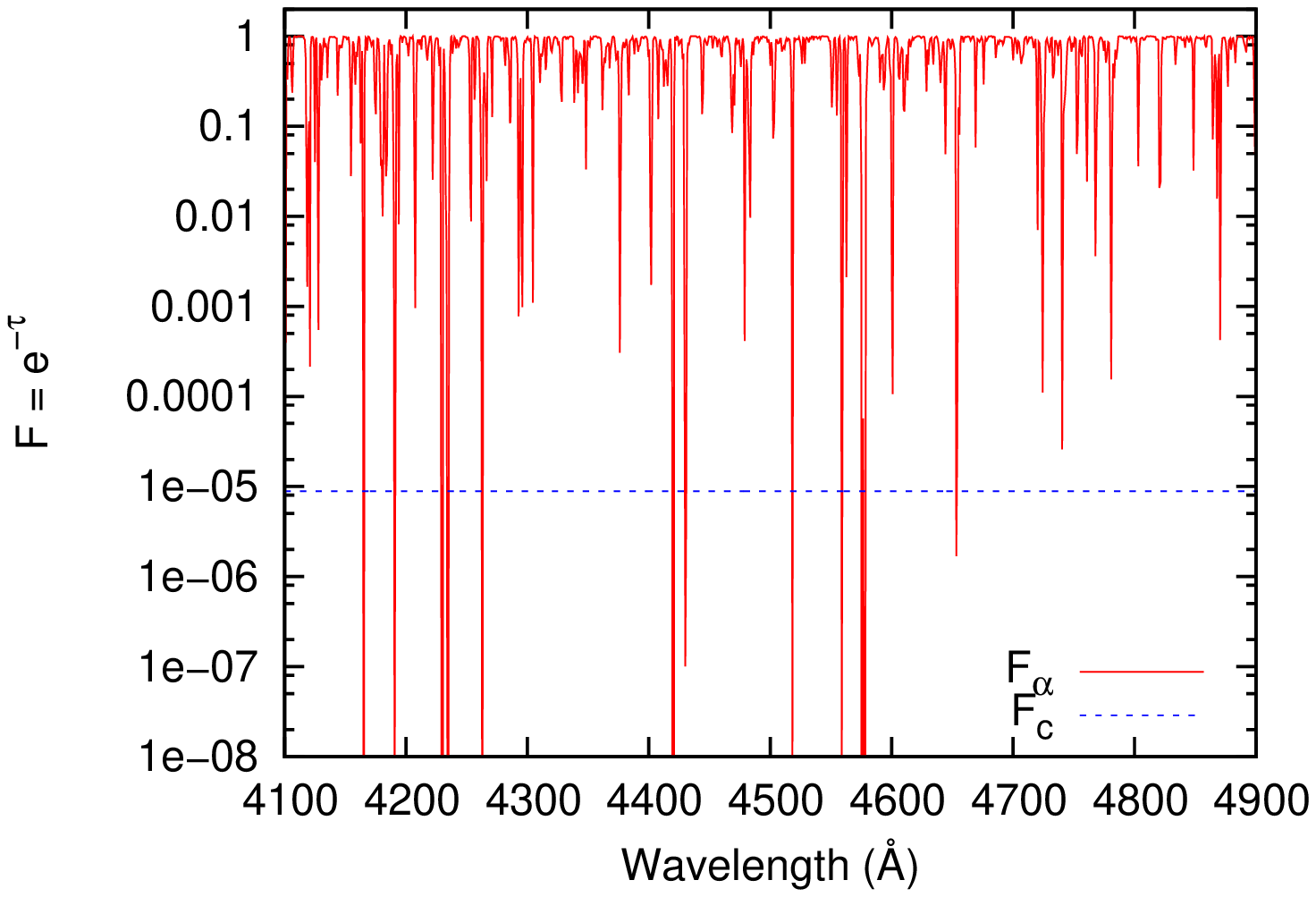}}
\subfigure{
	\includegraphics[scale=0.45, angle=-90]{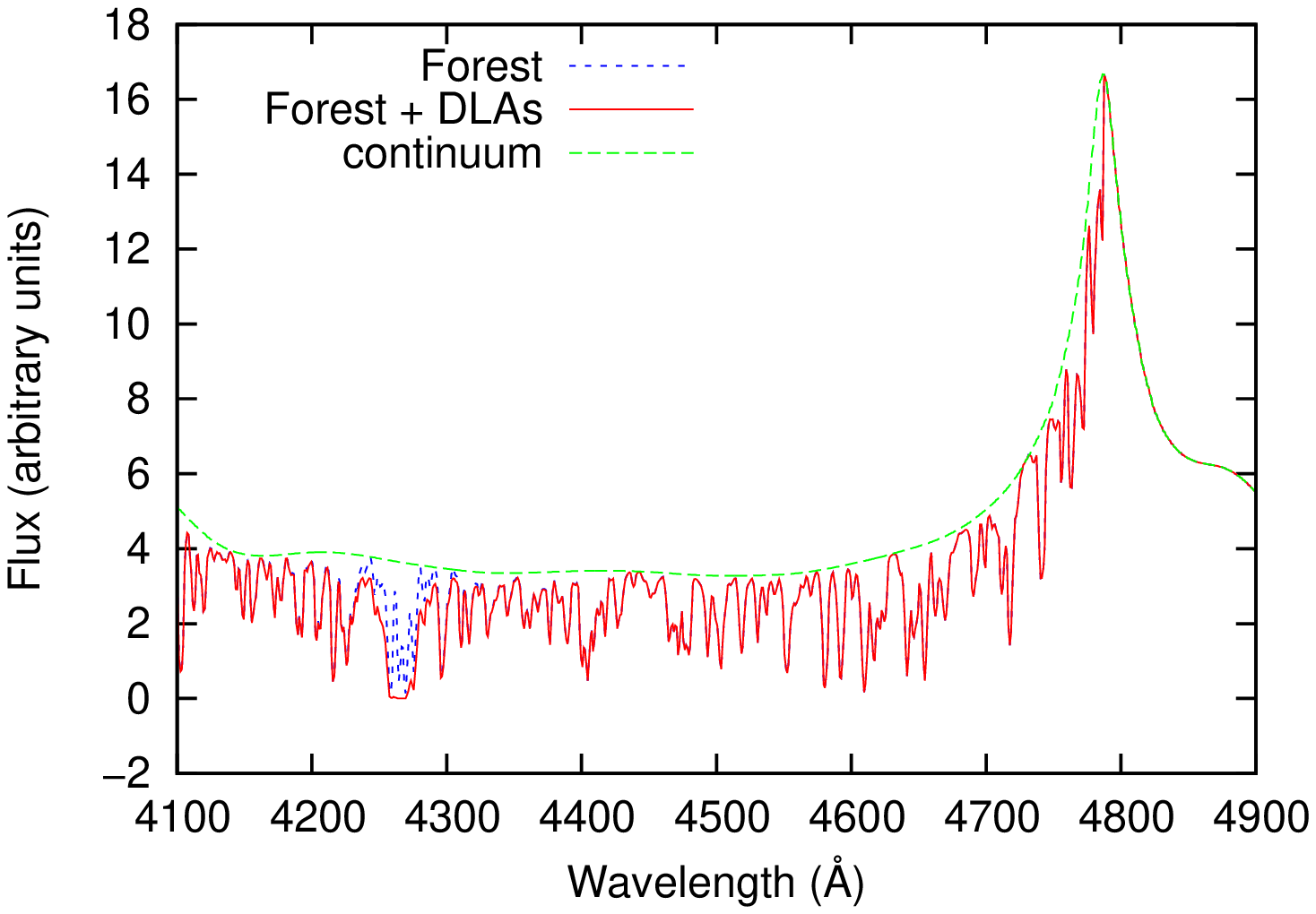}}
 \end{center}
 \caption{Left: Example of mock \lya absorption field $F$ (red) and the 
 threshold $F_c$ to host a HCD for $\nu=0.01$ (dotted blue). 
 Right: Mock spectrum for the same line of sight. The green line is the
 quasar continuum, and the blue line is the absorption due to \lya forest,
 smoothed with the spectrograph resolution as described in
 (\cite{2012JCAP...01..001F}). The red line includes absorption by a HCD.}
 \label{fig:example}
\end{figure}

  The HCD are randomly inserted in the mock spectra in pixels with
$\tau > \tau_c$, assigning column densities to them that follow the
distribution shown in Figure 1. A typical mock spectrum is shown in
Figure \ref{fig:example}. The left panel shows the spectrum of $F$ due
to the \lya forest and value of $F_c$. Systems can only be located in a
few narrow spikes in optical depth covering a fraction $\nu$ of the
spectrum. The right panel shows the spectrum in linear units after
multiplying by the quasar continuum and smoothing with the resolution
of the BOSS spectrograph, as described in \cite{2012JCAP...01..001F}. A HCD has
been randomly assigned to one of the peaks that cross the threshold in
the first figure (in this case, the peak at $\lambda \sim 4260$ \AA),
and has been included in the total absorption.

As shown in Appendix \ref{app:bias}, the large-scale clustering of the 
HCDs inserted in this way in the mock spectra follows linear theory
with a bias factor $b_h$ that depends on $\nu$ and on the probability
distribution function of $F$ (which is modelled as a lognormal function
in our mocks, as described in \cite{2012JCAP...01..001F}), and with a redshift
distortion parameter equal to that of the \lya forest,
$\beta_h=\beta_\alpha$. We note that this is a consequence of the
simple procedure we use for inserting the HCDs, and that in reality
their bias factor and redshift distortion parameter should depend on
the distribution of their host halos and the selection effects involved
in their detection.
For our fiducial value of $\nu=0.01$, the bias of the HCDs inserted
in our mocks is $b_h = 1.21$, while for a more extreme value of $\nu = 0.002$
the bias is $b_h = 1.43$.

%% file: Results.tex
\section{Effect on the measured \lya correlation function}
\label{sec:results}

  We now investigate the impact of HCDs on the \lya correlation
function, by measuring it directly in mock spectra.
We generate 100 realizations of a mock survey with an area of
$200 \, \rm{deg^2}$.
Quasars are distributed following the luminosity function from
\cite{2006AJ....131.2788J}, with a total quasar density of 
$22 \, \rm{deg^{-2}}$, over the redshift range $2.15 < z < 3.5$.
We use the same definition of the \lya forest as
\cite{2011JCAP...09..001S}, i.e., the rest-frame wavelength range
1041 \AA\ - 1185 \AA. We also apply a cut at an observed wavelength of
3600 \AA, close to the end of the BOSS spectrograph. 

%{\it From Jordi: Did you mean here that no \lya forest is
%generated for $z<2.15$? In the BOSS spectra, the \lya forest is used
%down to a lower redshift. Do you also consider the wavelength range
%from 1040 to 1185 angstroms?}
As previously mentioned, the \lya absorption mock spectra are generated
with no redshift evolution and assuming that the lines of sight are parallel.
HCD systems are inserted with the method explained above,
and we measure the correlation function in 150 linear bins in $r$ of
width $1 \hmpc$, and 20 linear bins in $\mu=\cos\theta$. The
correlation function in each bin $A$ is estimated by averaging over
all pixel pairs with a separation that is within the bin A,
\begin{equation}
  \hat{\xi}_A = \frac{\sum_{i,j \in A} \delta_{Fi} \delta_{Fj}}
{\sum_{i,j\in A} 1} ~,
\end{equation}
where the indices $i$ or $j$ label all pixels in the \lya forest region.
Here the weights are all set equal to unity because the mock spectra 
are noiseless.

\begin{figure}[h!]
 \begin{center}
   \includegraphics[scale=0.6, angle=-90]{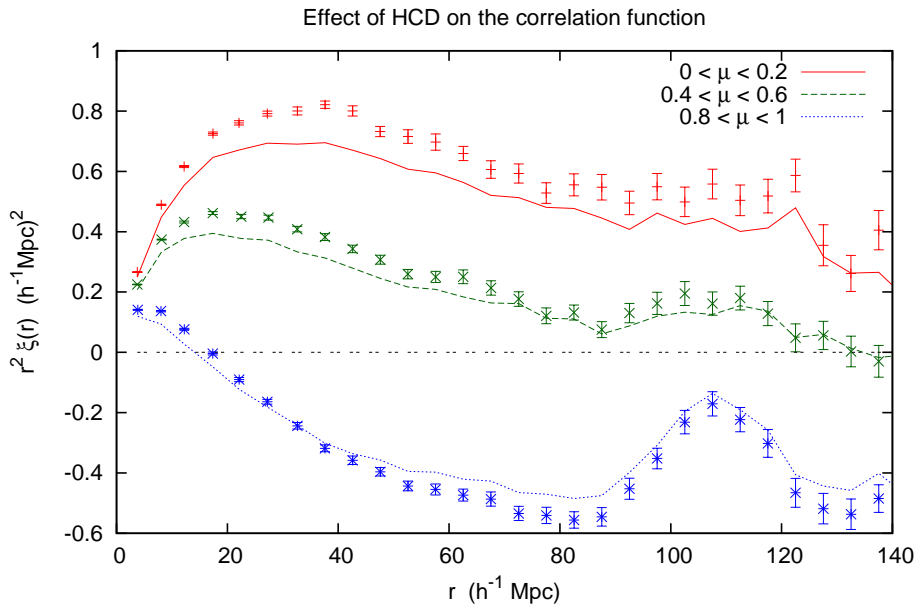}
 \end{center}
 \caption{Correlation function measured from mock spectra with inserted
  high column density systems, shown in 3 angular bins, with errorbars
  indicating the error for the mean of 100 mocks with 200 square degrees
  each. Thin lines show the mean measured in the same mock spectra
  without the HCD systems.
 } 
 \label{fig:dla_wedges}
\end{figure}

In Figure \ref{fig:dla_wedges} we show the mean measurement of the
correlation function in the $N=100$ realizations, before and after
adding the HCD systems, 
after averaging the original bins into wider ones with
$\Delta r = 5 \hmpc$ and $\Delta \mu = 0.2$. 
The errorbars are computed for the mean, equal to the dispersion among
realizations divided by $\sqrt{N-1}$.

  There are two main effects caused by the HCD systems on the measured
correlation function: the measurement error is increased, and the
correlation function is systematically altered from the true value
of the \lya forest alone. We first quantify the increase in the
statistical error in \S 3.1, and then we study the systematic effects
on the two \lya bias parameters in \S 3.2. We shall not examine in this
paper the systematic effect on the inferred shape of the linear power
spectrum in equation (2.1). However, as seen in Figure
\ref{fig:dla_wedges}, the systematic change in the correlation
function is not limited to the values of the bias parameters but can
affect the broadband shape in a generic way, even at very large scales
because of the extended damped wings of HCDs. Therefore, any other
constraints that are obtained from measurements of the \lya forest power
spectrum are in general subject to a correction for the impact of HCDs.

%The different panels in figure \ref{fig:dla_wedges} correspond to different
%models for the HCD absorbers. The first two panels show the effect when we
%introduce both LLS and DLAs, with a clustering parameter of $\nu=0.01$ 
%(top panel) and $\nu=0.002$ (bottom panel), corresponding to HCD bias of 
%XXX and YYY respectively. 

%\textcolor{red}{
%As expected, if only include LLS (bottom panel) the effect is reduced, but is 
%still significant in the linear regime, as will be quantified in the next 
%section. In none of the cases the BAO peak seems to be shifted, and the change 
%in the amplitude is just through the effective bias parameter.
%}

\subsection{Increase of measurement errors }

\begin{figure}[h!]
 \begin{center}
   \includegraphics[scale=0.6, angle=-90]{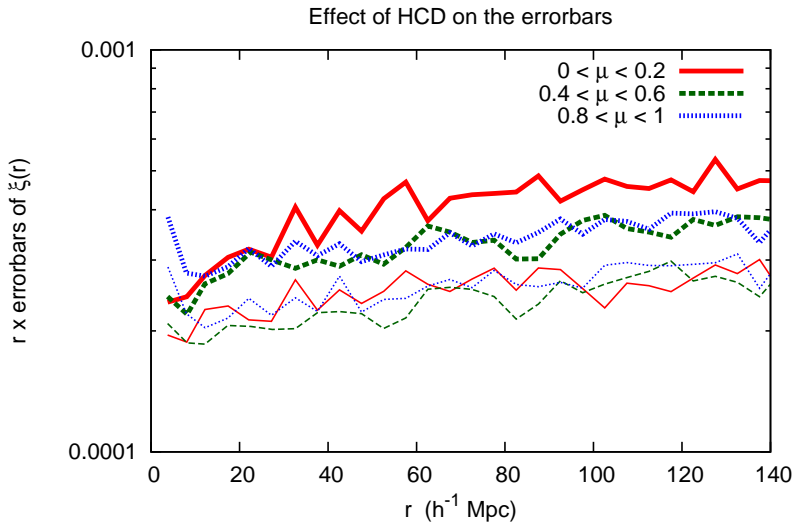}
 \end{center}
 \caption{Errors in the correlation function measured from mock spectra 
  with (thick lines) and without (thin lines) high column density systems,
  in 3 angular bins. The errors have been multiplied by $r$.
 } 
 \label{fig:errorbars}
\end{figure}

In Figure \ref{fig:errorbars}, the increase in the error of the measured
correlation function due to HCD systems is shown. The error is for the
mean of all the $N=100$ realizations.

The increase of the errorbars in the absence of observational noise
is $\sim 30 - 50\%$. Because errors are added quadratically, this
means that the contribution from HCD to the total noise is 
%nearly equal 
comparable to that arising from the intrinsic small-scale variance of the
\lya forest. In an actual survey
like BOSS, the error budget of the correlation function includes also
observational noise. For instance, it was shown in
\cite{2012JCAP...01..001F} that a level of noise comparable to that in the BOSS 
survey increases the errorbars by $\sim 30\%$. In this case, the total
error budget has three comparable contributions from the intrinsic \lya
forest variance, HCD systems and observational noise.

  We have performed some tests on the origin of the additional errors
introduced by HCDs in the correlation measurement. If DLAs (with
$N_{HI} > 10^{20.3} \cm^{-2}$ are eliminated from the HCDs that are
inserted (a model we designate as NO DLA, see Table 1 below), then the
increase of errors is reduced to 15\%. If the damped wings of all the
HCDs are eliminated (keeping only their saturated Gaussian profiles, a
model we designate as NO WINGS), then the error increase is further
reduced to 5 to 10\% of the total. This shows that the damped wings are
the dominant reason for increased errors in the correlation
measurements.

\subsection{Systematic effect on the measured bias parameters}

  We now discuss the result of fitting the correlation function in the
mocks with inserted HCD systems with the equation for the linear
theory power spectrum, 
\begin{equation}
 P_F(k,\mu_k) = b_F^2 \, (1+\beta_F \mu_k^2)^2 \, P_L(k) ~ ,
\end{equation}
where now $b_F$ and $\beta_F$ are the bias parameters for the total
absorption field, including \lya forest and HCD absorption.

  The precise procedure for fitting the correlation function by $\chi^2$
minimization carried out in \cite{2011JCAP...09..001S} requires a
complex and expensive calculation of the covariance matrix of the
correlation values measured in each pair of bins. Here we use a
simplified procedure, where only the diagonal elements of the covariance
matrix of the binned correlation function are taken into account to
minimize $\chi^2$. This allows us to quickly examine a large number of
realizations of many different models. However,
in order to obtain the errors in the fitted bias parameters, we rely
on bootstrap combinations of the 100 realizations of the survey. In
other words, we simply use the dispersion in the fitted values of the
parameters that are obtained in different realizations. Using the full
covariance matrix should lead to reduced errors of the fitted
parameters, but we find that this error reduction is not large and that
the systematic impact of HCDs is adequately reflected in the results
of the fits that use the diagonal elements of the covariance matrix
only.

  The results of the fits for the bias parameters are listed in Table
\ref{tab:fits} for a variety of models with different properties of the
HCDs. The mean effective optical depth and the bias of the selected
pixels for inserting HCDs are listed initially in Table 1. The
additional variable $C$ given in the Table is the correlation of the
\lya and HCD transmission fluctuation at zero separation [see eq.
(\ref{eq:cdef}) in the next section, where this will be used; we note
here that the mean transmitted fraction $\bar F$ is slightly reduced by
the insertion of HCDs according to eq. (\ref{eq:mfc}), where
$\bar F_H = \exp(-\bar\tau_{eH})$]. As shown
in \cite{2011JCAP...09..001S},
the combination of parameters that is most accurately determined from the
3D correlation function is $b_F \, (1+\beta_F)$. We therefore list the
average best fit values of $b_F$, $\beta_F$ and $b_F (1+\beta_F)$, each
one with the error directly obtained from the bootstrap analysis. We
note that the
linear function used to fit the correlation function in equation (3.2)
neglects the non-linear term $D(k,\mu_k)$ that is present in the power
spectrum used to generate the mocks (eq.\ 2.1). We therefore do not
expect to recover exactly the input values of the bias factors even
when no HCDs are included. To minimize the non-linear effects, our fit
to the correlation function uses only bins at $r > 10\, {\rm Mpc/h}$.

\begin{table}[h!]
 \vspace{5pt}
 \begin{center}
  \begin{tabular}{l|ccc|ccc}
  Model   & $\bar\tau_{eH}$ & $b_h$ & $C$ & $b_F$ & $\beta_F$ & $b_F (1+\beta_F)$ \\
  \hline
  NO HCD   	& 0 & -- & -- & $ -0.1472 \pm 0.0005 $ & $1.550 \pm 0.009$ & $-0.3756 \pm 0.0003$  \\
  FIDUCIAL   	& 0.017 & 1.21 & 0.0034 & $ -0.1678 \pm 0.0013 $ & $1.374 \pm 0.018$ & $-0.3984 \pm 0.0008$  \\
  HIGH BIAS  	& 0.016 & 1.43 & 0.0032 & $ -0.1732 \pm 0.0009 $ & $1.306 \pm 0.011$ & $-0.3994 \pm 0.0009$  \\
  NO DLA  	& 0.009 & 1.21 & 0.0029 & $ -0.1563 \pm 0.0006 $ & $1.495 \pm 0.011$ & $-0.3902 \pm 0.0004$  \\
  NO WINGS  	& 0.007 & 1.21 & 0.0033 & $ -0.1519 \pm 0.0006 $ & $1.546 \pm 0.010$ & $-0.3867 \pm 0.0005$  \\
  \end{tabular}
  \caption{Fitted bias parameters for mocks with different models for
the inserted HCDs. The variable $C$ is defined in the next section.}
  \label{tab:fits}
 \end{center}
\end{table}

  The first model in Table 1 (labeled NO HCD) does not include any HCD
systems. The recovered values of the bias parameters are very close to
the input ones, $b=-0.1375$ and $\beta=1.58$. The small differences are
due to the non-linear term. The second row is for our fiducial model,
where HCDs are added following the column density distribution described
in the previous section in regions of high optical depth with a spectral
filling factor $\nu=0.01$, corresponding to a bias factor $b_h=1.21$. 
The HCDs in the fiducial model induce an increase of the bias
parameter $b_F$ of 14\%, and a reduction of the redshift distortion
parameter $\beta_F$ of 11\%.

  The third model, labeled HIGH BIAS, forces the HCDs into a smaller
fraction of the \lya forest spectra, $\nu=0.002$, increasing their bias
factor to $b_h=1.43$.
The value of $b_F$ now increases by 18\% and $\beta_F$ decreases by
16\%. The systematic impact of HCDs is therefore increased as their
bias factor increases. We remind that here we have to keep the value of
$\beta_h$ for the HCDs equal to that of the \lya forest, because of the
way they are inserted in the mock spectra. In reality $\beta_h$
should be close to $1/b_h$ and, as we shall see below this should
further enhance the impact of the HCDs on the value of $\beta_H$.

  The fourth row gives the result for a model where only systems with
a column density $N_{H_I} < 10^{20.3} \cm^{-2}$ are included (labeled
NO DLA), using again $\nu=0.01$. In other words, the systems generally
referred to as damped \lya systems are not included. The lower column
density systems producing the remaining effect, even though they are not
generally identified as DLAs, obviously produce weak damped absorption
wings as well. These weak systems are responsible for an increase of the
bias factor of 6\% and a decrease of $\beta_F$ of 3.5\%, i.e., $\sim$
30 to 40\% of the total effect of all the systems in the
FIDUCIAL model. In a survey with spectra with the resolution and
signal-to-noise of BOSS, most of the systems with $N_{H_I} > 10^{20.3}
\cm^{-2}$ can be individually identified and removed from the sample in
order to test for their impact on the correlation function, but most of
the systems with lower column densities cannot be reliably identified.
Therefore, removing the identified DLAs from the sample will not
completely eliminate the systematic errors in the \lya correlation
induced by HCDs. We note here that if one chooses to eliminate all the
spectra containing DLAs in a \lya forest survey like BOSS, then the
measured correlation is also systematically biased in a different way
because the \lya forest is correlated with the presence of HCDs.

  The final model (fifth row, labeled NO WINGS) inserts all the HCDs
only with their Gaussian profiles (with the fixed Doppler parameter
$b_D= 70 \kms$), with no damped wings. In this case the impact on the
recovered bias parameters is smaller, especially for $\beta_F$ which
is practically not affected. This shows that the main impact of HCDs
on the correlation function is through the damped wings of the
absorbers.

%% file: Effect_Dlas.tex
\section{Analytical description}
\label{sec:effectHCD}

  We now present an analytical formulation to evaluate the effect of
the high column density systems (HCDs) on the correlation function of 
the flux transmission fraction, $F$. Even when the analytical results
require making certain approximations, they are highly useful to provide
an interpretation of the numerical results and an understanding of the
dependence to be expected with any variations of the model for the
column density distribution and bias parameters of the HCDs.

  We start by introducing some useful notation. The transmitted fraction
at a point $\vx$ in a spectrum is
$F(\vx)= \bar F \left[1+\delta_F(\vx)\right]$, where $\bar F$ is the
mean value of $F$ at a certain redshift. We divide this total
transmission into a contribution $F_H$ from HCDs, defined as absorption
systems with a column density $N_{HI} > 1.6\times 10^{17}\cm^{-2}$
(i.e., a continuum optical depth greater than unity at the Lyman limit),
and a contribution $F_\alpha$ from the \lya forest, defined as all the
remaining \lya absorption by atomic hydrogen. Hence,
 $F(\vx)=F_H(\vx)\, F_\alpha(\vx)$.
We ignore here the presence of metal lines; these will be considered
briefly in \S 5. We note that the precise column density at which this
conventional separation between \lya forest and HCDs is made does not
affect our results. The important point is that the \lya forest
absorption is dominated by systems with much lower column density than
the Lyman limit threshold, and the HCDs absorption is dominated by
systems with much higher column density than this threshold. Therefore,
the precise choice for the threshold is not crucial.

  Let the transmitted fraction of the \lya forest be
$F_{\alpha}(\vx) = \bar F_{\alpha} \left[1+\delta_{\alpha}(\vx)\right]$,
and the transmitted fraction of the HCDs be $F_H(\vx)= \bar F_H
\left[1+\delta_H(\vx)\right]$. We then have,
\begin{equation}
F(\vx) = \bar{F} \left[ (1 + \delta_F(\vx) \right] =
 F_\alpha(\vx) \, F_H(\vx) =
          \bar F_\alpha \left[1+\delta_\alpha(\vx)\right]
          \bar F_H \left[1+\delta_H(\vx)\right] ~.
\end{equation}
Being tracers of the same underlying density field, the fields
$\delta_\alpha$ and $\delta_H$ are correlated,
\begin{equation}
 C \equiv \, \left< \,\delta_\alpha(\vx) \delta_H(\vx) \,\right> \, \neq 0 ~,
\label{eq:cdef}
\end{equation}
and the relation between $\bar{F}$ and $\bar F_\alpha$ is
\begin{equation}
 \bar{F} = \, \left< F \right> \, = \bar F_\alpha \bar F_H (1+C) ~.
\label{eq:mfc}
\end{equation}
Hence, the variable $\delta_F(\vx)$ can be expressed as
\begin{equation}
 1 + \delta_F(\vx) = \frac{F(\vx)}{\bar{F}}
              = \frac{\left[1+\delta_\alpha(\vx)\right]\,
                      \left[1+\delta_H(\vx)\right]}{1+C} ~.
\end{equation}

\subsection{Impact on the correlation function}

 We can now write an expression for the correlation function of
$\delta_F$ at two points $\vxone$ and $\vxtwo$ as a function of the
separation $\vRot = \vxone - \vxtwo$:
\begin{align}
 \label{eq:lyadla}
   \nonumber
1 + \xi_F(\vRot) = & \left< \,\left[1+\delta_F(\vxone)\right]\left[1+\delta_F(\vxtwo)\right]\, \right> \\
   \nonumber = &\left( 1+C \right)^{-2} 
   \left< \,\left[1+\delta_\alpha(\vxone)\right] \left[1+\delta_H(\vxone)\right]
   \left[(1+\delta_\alpha(\vxtwo)\right] \left[(1+\delta_H(\vxtwo)\right] \,\right> \\
   \nonumber = &\left( 1+C \right)^{-2} 
                [ 1 + 2C + \xi_\alpha (\vRot) 
                    + 2 \xi_{\alpha H} (\vRot) + \xi_H (\vRot) \\
               &      + 2 \xi_{3\alpha} (\vRot)  
                    + 2 \xi_{3H} (\vRot)  
                    + \xi_{4} (\vRot) ] ~,
\end{align}
where we have defined:
\begin{align}
 \label{eq:termdef}
\nonumber \xi_{\alpha}(\vRot) &= \,
             \left< \,\delta_\alpha(\vxone) \delta_\alpha(\vxtwo)\, \right> ~, \\
\nonumber \xi_{\alpha H}(\vRot) &= \,
             \left< \,\delta_\alpha(\vxone) \delta_H(\vxtwo)\,\right> ~, \\
\nonumber \xi_H(\vRot) &= \,
             \left< \,\delta_H(\vxone) \delta_H(\vxtwo)\, \right> ~, \\
\nonumber \xi_{3\alpha}(\vRot) &= \,
         \left< \,\delta_\alpha(\vxone) \delta_H(\vxone) 
             \delta_\alpha(\vxtwo)\,\right> ~, \\
\nonumber \xi_{3H}(\vRot) &= \,
         \left< \,\delta_\alpha(\vxone) \delta_H(\vxone) 
              \delta_H(\vxtwo)\,\right> ~, \\
           \xi_{4}(\vRot) &= \,
         \left< \,\delta_\alpha(\vxone) \delta_H(\vxone) 
              \delta_\alpha(\vxtwo) \delta_H(\vxtwo)\,\right> ~.
\end{align}

\begin{figure}[h!]
 \begin{center}
  \subfigure{\includegraphics[scale=0.6, angle=-90]{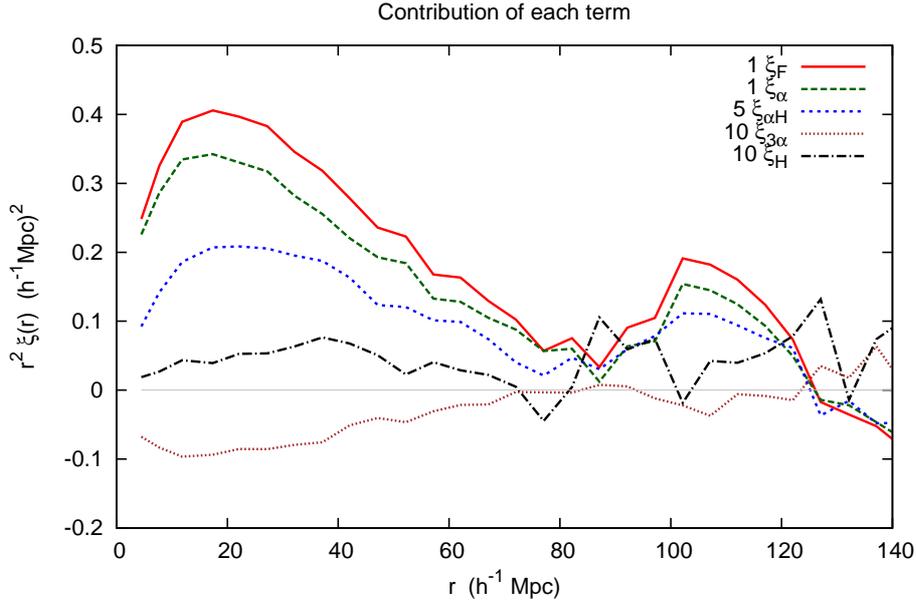}}
 \end{center}
 \caption{The total correlation function $\xi_F$ and the contribution
of the four largest terms in equation (4.6) in the numerical mock spectra.}
 \label{fig:terms}
\end{figure}

  We can compute the contribution of each of the 6 terms in equation
\ref{eq:lyadla} for the case of the mock spectra discussed in Sections
2 and 3. Figure \ref{fig:terms} shows these contributions for our
FIDUCIAL model, for the largest four terms only. The remaining two terms
are substantially smaller and are omitted, and some of the
terms are shown multiplied by a factor of 5
or 10 for better visualization.
As seen in this figure, the leading
correction due to the presence of HCDs is the term $\xi_{\alpha H}$,
although the next largest term, $\xi_{3\alpha}$, is not much smaller.

\subsection{Effective bias parameters}

  The full expression for the overall transmitted fraction correlation
in equation (\ref{eq:lyadla}) is highly complex, depending on a
combination of 2, 3 and 4-point functions of the fields. While the
3-point and 4-point functions defined in equation (\ref{eq:termdef}) are
difficult to characterize, the two-point correlation terms can be
calculated in the limit of large scales assuming that linear theory
applies. Then, each Fourier mode is multiplied by the factor
$1+\beta_i \mu_k^2$, with $\beta_i$ being the redshift distortion factor
of any tracer $i$ \cite{1987MNRAS.227....1K}. With this in mind, we
first aim to solve analytically the case when the 3-point and 4-point
functions are negligible. Note, however, that in the case of our
FIDUCIAL model in the numerical mocks, the 3-point term $\xi_{3\alpha}$
is actually larger than $\xi_H$ and is smaller than $\xi_{\alpha H}$ by
a factor of only $\sim$ 5, as shown in Figure 6. Furthermore, assuming
linear theory is not actually accurate in this case, because the damped
wings of the absorbers extend the correlation in the radial direction up
to scales that can be arbitrarily large depending on the shape of the
correlation function. The results obtained by considering only the
linear 2-point terms should therefore be considered as no more than an
initial guide, to be revised by the effects of the other terms (in
particular $\xi_{3\alpha}$ for the observationally more relevant cases)
and the validity of the linear approximation.

  We therefore separate equation \ref{eq:lyadla} as
\begin{equation}
 \xi_F(\vRot) = \xi_2(\vRot) + \xi_{34}(\vRot) ~,
\end{equation}
where the 2-point contribution is
\begin{equation}
 \xi_2 (\vRot) \equiv \frac{\xi_{\alpha} (\vRot) 
                    + 2 \xi_{\alpha H} (\vRot) + \xi_H (\vRot)}
                    {\left( 1+C \right)^2 } ~,
\label{eq:xi2def}
\end{equation} 
and all the remaining 3-point and 4-point terms are included in
$\xi_{34}$,
\begin{equation}
 \xi_{34} (\vRot) \equiv \frac{ 2 \xi_{3\alpha} (\vRot) 
             + 2 \xi_{3H} (\vRot)
             + \xi_{4} (\vRot) - C^2}{ (1+C)^2} ~.
\label{eq:xi34}
\end{equation}
Note that in the limit of large separation, $\xi_4$ approaches
$C^2$ and therefore $\xi_{34}$ vanishes.
 
  Now, the Fourier transforms of the 2-point correlations $\xi_\alpha$,
$\xi_H$ and $\xi_{\alpha H}$ yield their corresponding power spectra,
which can be expressed in terms of the bias factors for each field. As
described in \cite{2000ApJ...543....1M} and \cite{2003ApJ...585...34M},
the general bias parameters of an absorption field arise by considering
the first order expansion of the average of the mean transmission in a
large scale, linear region that is conditioned to have density and
peculiar velocity gradient perturbations $\delta$ and $\eta$.
Therefore, in analogy to the \lya forest, the bias factors for the HCD
absorption field $\delta_H$ are
\begin{equation}
  b_H = {1 \over \bar F_H}{\partial F_H \over \partial \delta} \, ,
\qquad
  b_{\eta H} = {1 \over \bar F_H}{\partial F_H \over \partial \eta} ~, 
\label{eq:biash}
\end{equation}
and the redshift distortion parameter is
\begin{equation}
  \beta_H = { b_{\eta H} f(\Omega) \over b_H} ~,
\label{eq:betah}
\end{equation}
where $f(\Omega)$ is the logarithmic derivative of the growth factor
(see \cite{1987MNRAS.227....1K}).
In linear theory, each Fourier mode of any tracer field is multiplied by
the factor $b_i (1+\beta_i \mu_k^2)$ relative to the mass field, and
therefore the power spectra are
\begin{equation}
 P_{\alpha}(k,\mu_k) = b_\alpha^2 (1+\beta_\alpha \mu_k^2)^2 P_L(k) ~,
\label{eq:powa}
\end{equation}
\begin{equation}
 P_H(k,\mu_k) = b_H^2 (1+\beta_H \mu_k^2)^2 P_L(k) ~,
\label{eq:powh}
\end{equation}
\begin{equation}
 P_{\alpha H}(k,\mu_k) = b_\alpha b_H\, (1+\beta_\alpha \mu_k^2) 
                       (1+\beta_H \mu_k^2)\, P_L(k)  ~.
\label{eq:powah}
\end{equation}

  Finally, the power spectrum $P_2$ of the 2-point terms is
\begin{align}
 \nonumber \frac{P_2(k,\mu_k)}{P_L(k)} &= \frac{
    b_\alpha^2 (1+\beta_\alpha \mu_k^2)^2 
    + 2 b_\alpha b_H (1+\beta_\alpha \mu_k^2) (1+\beta_H \mu_k^2) 
    + b_H^2 (1+\beta_H \mu_k^2)^2}{\left( 1+C \right)^2} \\
 &= b_2^2 (1+\beta_2 \mu_k^2)^2 ~,
\label{eq:ptwo}
\end{align}
where we have defined
\begin{equation}
 \label{eq:bias_eff}
 b_2 \equiv \frac{b_\alpha + b_H}{1+C} ~,
\end{equation}
and
\begin{equation}
 \label{eq:beta_eff}
 b_2 \beta_2 \equiv \frac{ b_\alpha \beta_\alpha + b_H \beta_H} {1+C} ~.
\end{equation}

  In the absence of the term $\xi_{34}$ and for linear theory, the
relation between the bias factors measured from the observed correlation
and that of the unpolluted \lya forest is therefore remarkably simple.
In principle, both $b_H$ and $\beta_H$ can be measured from the
observable cross-correlation of the HCDs with the \lya forest, and then
the systematic effect of the HCDs on the total correlation can be
corrected to obtain the \lya forest bias parameters.

\subsection{Relation to the bias of host halos}
\label{ss:halobias}

  Whereas most of the \lya forest absorption at $z>2$ is associated with
density fluctuations in the intergalactic medium forming an
interconnected structure, the high column density systems should
correspond to discrete, clearly identifiable overdense regions that
have gravitationally collapsed, or halos. The natural question that
arises is the relation of the bias factors defined in equation
(\ref{eq:biash}) to the host halo bias. We now address this question
in order to predict the bias factors and the impact of the HCDs on the
total power spectrum from equation (\ref{eq:ptwo}).

%Note that there is always some
%remaining ambiguity in the identification of halos as separate objects
%when the halos are in the process of merger events, 
%but only a small fraction of
%halos are undergoing a merger at any given time. The question that
%arises then is the relation between the bias factor of the halos hosting
%the HCDs and the bias factor of the HCDs when measured in the absorption
%spectra. This
%relation is in general complicated because the HCDs in absorption are
%clustered and their absorption profiles in the spectra can be blended in
%a non-linear way, in which their absorption equivalent widths are not
%simply added up. However, 

  We define new bias factors for the HCDs, which can be defined in an
equivalent way for any other population of absorbers,
\begin{equation}
  b_H' = {1\over \bar \tau_{eH}}{\partial \tau_{eH}\over \partial\delta} =
  - {b_H\over \bar\tau_{eH}}\, ; \qquad 
  b_{\eta H}' = - {b_{\eta H}\over \bar\tau_{eH}} ~,
\end{equation}
where $\tau_{eH} = -\log F_H$ is the {\it effective} optical depth
averaged over a large scale region with mean values of the density and
peculiar velocity gradient perturbations $\delta$ and $\eta$, and
$\bar\tau_{eH}$ is its average over all the universe, for
$\delta=\eta=0$. The effective optical depth is obtained by averaging
$F_H$ first, and then taking the logarithm. The bias is defined
analogously to equation (\ref{eq:biash}), but using the effective
optical depth instead. These new bias factors for an absorption field
can be interpreted in the usual way that bias factors for a collection
of objects are interpreted: if the mean density perturbation increases
by a fractional amount $\delta$, while $\eta$ is kept fixed, the mean
effective optical depth of HCD absorption increases by a fractional
amount $b_H' \delta$.

  If the bias factor of the HCD host halos is $b_h$, this means that
their number density should fluctuate on large scales as $\delta_h = b_h
\delta$. If we now assume that the probability of observing an HCD when
a halo of fixed mass is intercepted is independent of its large-scale
environment (i.e., independent of $\delta$ and $\eta$), then the
perturbation in the effective optical depth contributed by HCDs should
be the same as that in the halo number density. In other words, we
should have $b_H'= b_h$, and $b_{\eta H}'=1$ in redshift space. More
generally, this assumption holds only if the following two conditions
are met:

\begin{enumerate}
 \item The probability that the absorption profile of any HCD appears
substantially blended with another one in the absorption spectrum is
small. Here, substantially blended means that their profiles overlap in
a region where their absorption optical depth is not much smaller than
unity. This condition should in general be correct if
$\bar\tau_{eH} \ll 1$ and the clustering of HCDs is not very strong.

 \item The probability distribution of the column density in a halo
of a fixed mass $M_h$ is independent of its large-scale environment and
is isotropic (i.e., it is independent of $\delta$ and $\eta$). In other
words, the gas radial profile does not depend on the environment for
fixed $M_h$, and the axes of any non-spherical gas distribution in the
halos are not correlated with the principal axes of the deformation
tensor of the surrounding large-scale structure. This assumption is
likely to be not exactly true, because galaxy disks are
known to be statistically aligned with the axes of their large-scale
environment, which can affect their redshift distortion anisotropy
\citep{2009MNRAS.399.1074H}, but the effect is probably very small.
\end{enumerate}
Under these conditions, the transmission fluctuations due to HCDs
should obey $\delta_H ({\bf x}) = - \bar\tau_{eH} \delta_h({\bf x})$,
and the bias factors are related by
\begin{equation}
 b_H = - \bar\tau_{eH} b_H' = -\bar\tau_{eH} b_h ~; \qquad\qquad
 \beta_H = {b_{\eta H} f(\Omega) \over b_H } = {f(\Omega)\over b_h} ~.
\label{eq:bhrel}
\end{equation}
In the rest of this Section, we assume that these relations hold,
which is reasonable for HCDs in view of the conditions that are
required, and that $\bar\tau_{eH} \ll 1$. Note that these new bias
factors can be defined for the \lya forest in the same way, and that
they also provide a measure of the relative fluctuations in the \lya
effective optical depth in comparison to the relative fluctuations in
the mass density, but the \lya forest absorption cannot be associated
with halos and the two assumptions above are not correct for this case.

\subsection{Corrections for the two-point linear terms}

Using the results above, we can now derive the correction to the bias
factors measured from the total absorption field that includes the \lya
forest and the HCDs, to obtain the corrected bias factors of the \lya
forest: 
\begin{equation}
 \label{eq:db1}
 \Delta b \equiv b_2 - b_\alpha = \frac{b_\alpha + b_H}{1+C} - b_\alpha 
            = - {\bar\tau_{eH} b_h + C b_\alpha \over 1+C }
            \simeq - \bar\tau_{eH} b_h ~,
\end{equation}
\begin{equation}
 \label{eq:db2}
 \Delta \left(b \beta\right) \equiv b_2 \beta_2 - b_\alpha \beta_\alpha 
    = - { \bar\tau_{eH} f(\Omega) + b_\alpha \beta_\alpha C \over 1+C} 
   \simeq - \bar\tau_{eH} f(\Omega) ~ ,
\end{equation}
\begin{equation}
 \label{eq:db3}
 \Delta \beta \equiv \beta_2 - \beta_\alpha
    = {b_\alpha \beta_\alpha - \bar\tau_{eH} f(\Omega) \over b_2 (1+C) }
      - \beta_\alpha
    = {\bar\tau_{eH} \left[ \beta_\alpha - f(\Omega)/b_h \right] \over
       b_\alpha/b_h - \bar\tau_{eH} } ~,
\end{equation}
where the last expressions in equations (\ref{eq:db1}) and
(\ref{eq:db2}) assume $C \ll 1$ ($C$ is usually also substantially
smaller than $\bar\tau_{eH}$ if this effective optical depth from
HCDs is dominated by damped wings, which have little correlation with
the \lya forest). In Table \ref{tab:fits} we show the value of $C$ for the
different mocks. Finally, the correction on the parameter that is best
constrained from the 3D power spectrum is
\begin{equation}
 \Delta \left(b + b \beta \right) = - { \bar\tau_{eH}
   \left[ b_h + f(\Omega) \right] + C b_\alpha (1+\beta_\alpha) \over
       1+C } \simeq - \bar\tau_{eH} \left[ b_h + f(\Omega) \right] ~.
 \label{eq:db4}
\end{equation}

\begin{figure}[h!]
 \begin{center}
   \includegraphics[scale=0.5, angle=-90]{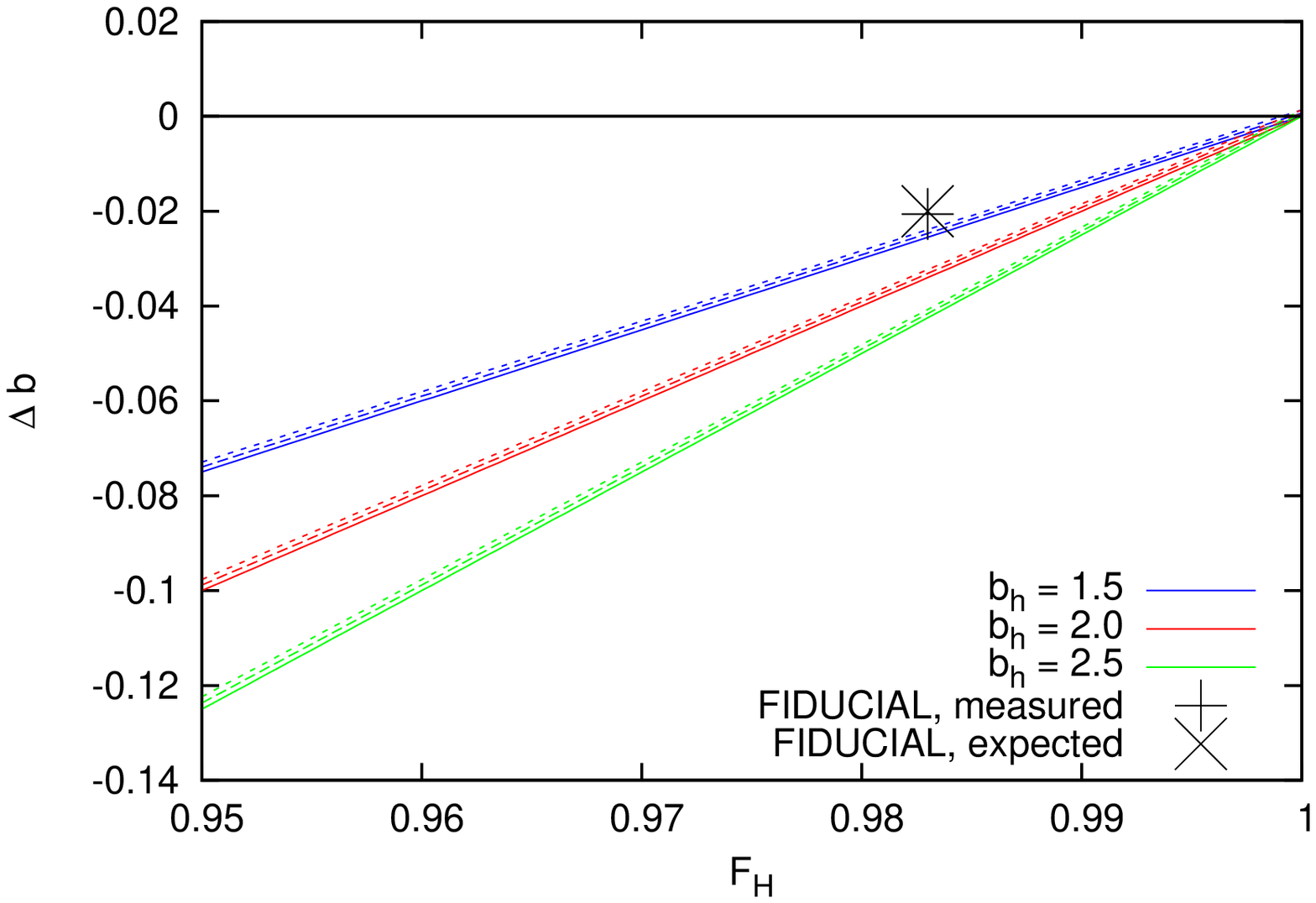}
   \includegraphics[scale=0.5, angle=-90]{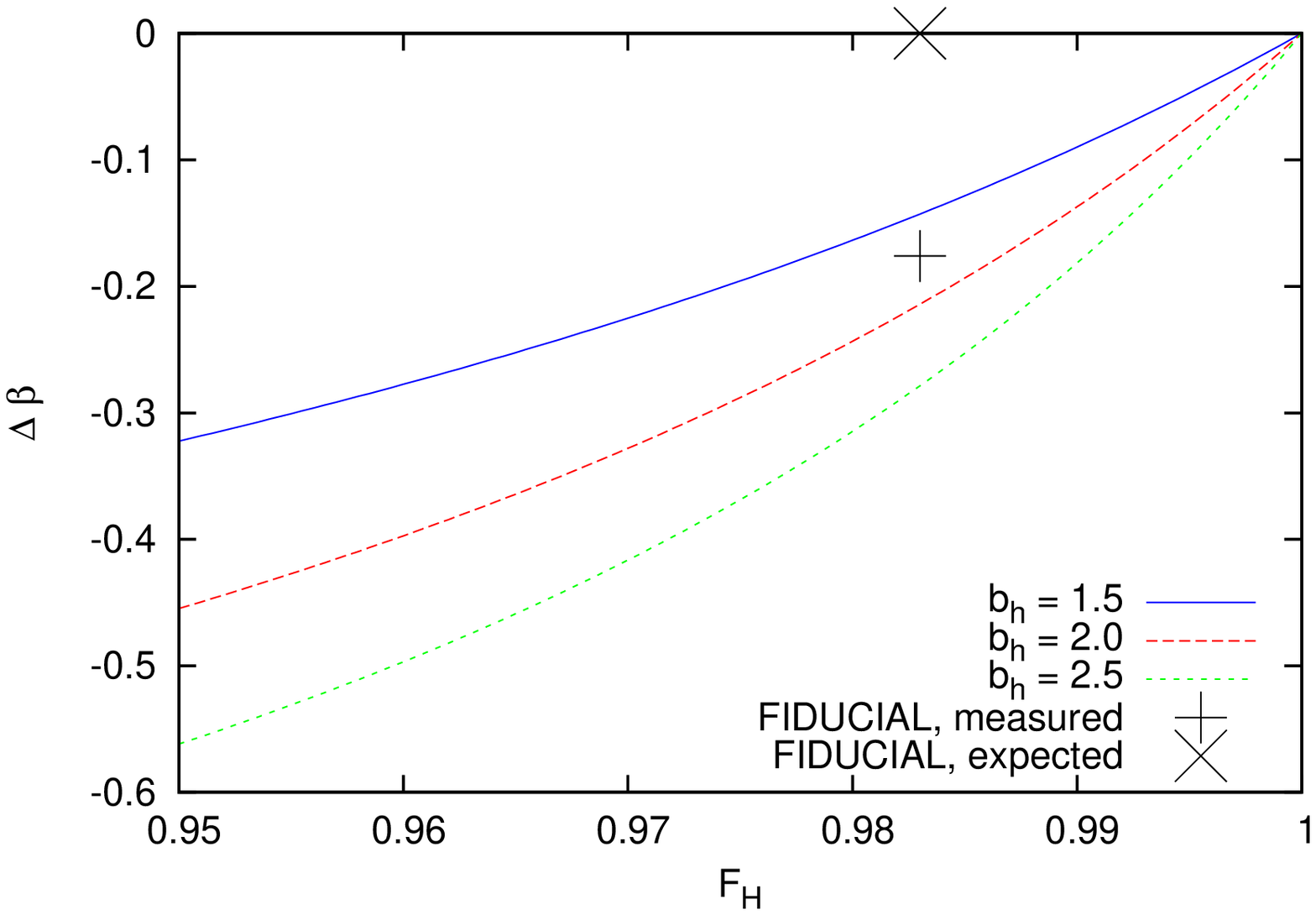}
   \includegraphics[scale=0.5, angle=-90]{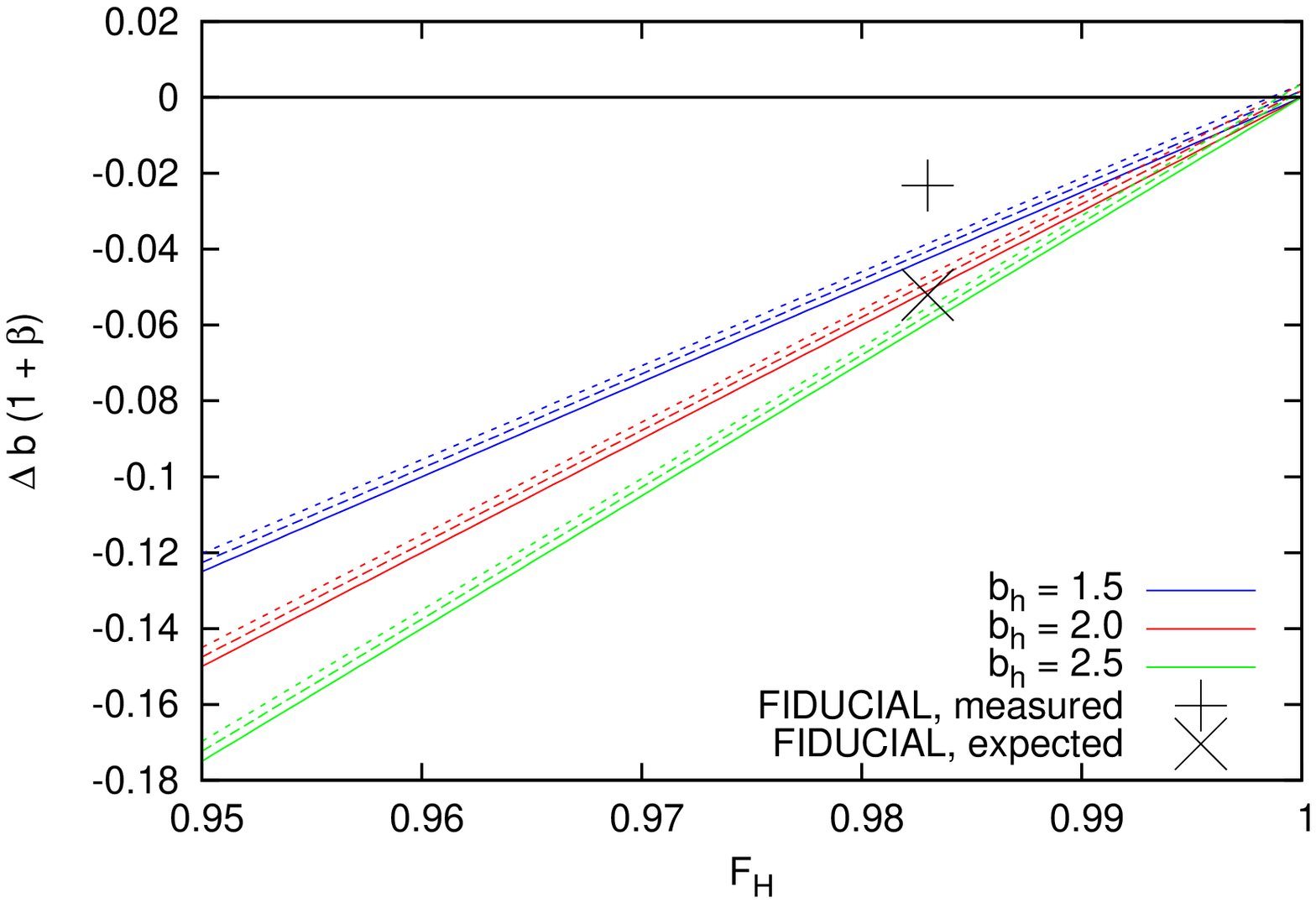}
 \end{center}
 \caption{Correction to bias parameters as a function of halo bias $b_h$. 
  Solid, dashed and dotted lines correspond to different values of $C$ 
  (0,0.01,0.005). Crosses show the analytical prediction for the
  FIDUCIAL mocks, and the actual measurement from the mocks.}
 \label{fig:eff_bias}
\end{figure}

  The simple bias parameter corrections in equations \ref{eq:db1},
\ref{eq:db2}, \ref{eq:db3} and \ref{eq:db4} are
plotted in Figure \ref{fig:eff_bias} as a function of $\bar\tau_{eH}$,
for several values of $b_h$, and for the \lya forest bias values used in
our numerical mocks, $b_\alpha=-0.1375$ and $\beta_\alpha = 1.58$.

  The effect of the HCDs can be mitigated if a large fraction of them
can be individually detected in the absorption spectra, when the
signal-to-noise is good enough. In a survey with similar
characteristics as BOSS, one should be able to detect and mask most of
the DLAs (with $N_{HI} > 10^{20.3}\cm^{-2}$), considerably reducing the
value of $\bar\tau_{eH}$. The lower lines in Figure \ref{fig:FH} show
$1-\bar F_H \simeq \bar\tau_{eH}$ when only systems of lower column
densities are left. The effective optical depth from these systems is
still about one third of that of all the HCDs.

  However, masking and removing some of the HCDs before the flux
correlation is measured may introduce potential problems. The HCDs are
correlated with the \lya forest, so if a portion of the spectrum around
each HCD is simply eliminated from the data, we are introducing a bias
that may be comparable or worse than the one caused by the HCDs
themselves. Moreover, the HCDs that are detected may suffer from
selection effects induced by the superposition of their damped wings
with the \lya forest. For example, HCDs living in large-scale regions
with different values of $\eta$ may have different probabilities of
being detected because the \lya forest absorption around them depends on
$\eta$, and this would change the derived values of the $b_\eta$ and
$\beta$ parameters. One should therefore be particularly careful if
DLAs are masked and removed to precisely simulate the effect of this
procedure in numerical mocks and see if the noise they introduce in the
measurements can in fact be substantially reduced while any systematic
effects that their removal may induce are properly corrected for.

\subsection{Application to the measurement on mock spectra}

Our simple analytical estimate for the systematic effect of HCDs on the
bias parameters can now be compared to the numerical results obtained
previously from mock spectra in the last section. Note that in our
numerical mocks, $\beta_H = \beta_\alpha$ and 
$b_{\eta H}' = b_h \beta_\alpha / f(\Omega) \neq 1$, so
equations (\ref{eq:db1}), (\ref{eq:db2}), (\ref{eq:db3}) and
(\ref{eq:db4}) cannot be used, or must be modified to include the
correct $b_{\eta H}$. Instead, we can use directly equations
(\ref{eq:bias_eff}) and (\ref{eq:beta_eff}). The parameters for our
fiducial mocks are $b_h=1.21$, $\beta_h=\beta_\alpha=1.58$, $C=0.0034$, 
$b_\alpha=-0.1375$ and $\bar\tau_{eH}=0.017$, which yield
\begin{equation}
 \Delta b = - 0.02, \qquad \Delta \beta = 0 
\qquad \Delta b(1+\beta) = -0.052 ~ .
\end{equation}
These corrections from the simple two-point, linear model are shown
as crosses in Figure 7. Clearly, other effects must be important in
changing the best fit bias factors in the mocks, in which the value of
$\beta$ is substantial decreased by the HCDs.

% and a smaller decrease of $b$ than our simple analytical prediction.

  There are two reasons for the discrepancies between the analytically
predicted corrections to the bias parameters and the actual corrections
found in the mocks. One is the presence of the 3-point term
$\xi_{3\alpha}$, the dominant one among the terms in $\xi_{34}$
(eq. \ref{eq:xi34}) that are not taken into account. However,
Figure \ref{fig:terms} shows that this term contributes only $\sim 2\%$
to the total correlation function, and is $\sim$ 5 times smaller than
the two-point terms, so it is surprising that its impact on the fitted
values of the bias parameters of the numerically obtained correlation
function may be so large.

  A clue to what is going on is found in Table 1, where the model
NO WINGS (in which the damped wings of the HCDs are eliminated) shows
practically no correction on $\beta_F$. This model has a value of
$\bar\tau_{eH}=0.0068$,
so we would expect the correction on $\beta_F$ to be about three times
smaller than for the FIDUCIAL model under no other changes, but the
change on $\beta_F$ is actually 30 times smaller.
%The change in the
%value of $b_F$ in the NO WINGS model is also in much better agreement
%with our analytical estimates above than in the case of the FIDUCIAL
%model. 
This strongly suggests that the reason for the discrepancy is to be
found in the effect of the damped wings.

\subsection{The non-linear effect of the damped wings}

  The damped wings imply that the linear theory approximation is not
actually valid even on very large scales. The damped absorption falls
as the inverse square of the line-of-sight separation, a dependence
that is comparable to the rate of decrease of the correlation function.
Therefore, the actual correlation function that is observed is not
the linear theory expression derived from the power spectra in
equation (\ref{eq:ptwo}), but is the convolution along the line of
sight of this linear
correlation function with the Voigt profile of the average column
density in HCDs that is correlated with the \lya forest absorption.

  Quantitatively, the linear form of the cross-correlation
$\xi_{\alpha H}(\vxperp,v)$, where the parallel component of the
separation $v$ is assumed to be expressed as a velocity, assumes that
the absorption of a HCD is localized into an interval much narrower than
the separation $v$. The damped wings imply that this is not actually
true, and that the absorption has the Voigt profile $V(N,v'-v)$ at a
velocity $v'$, for an absorber located at $v$ with a column density
$N$. Then, if the probability distribution of $N$ is proportional to
$f(N)$, and neglecting any effects of blending of the HCD profiles among
themselves, the cross-correlation of the \lya and HCD absorption is
modified to a function $\xi^V_{\alpha H}$ equal to
\begin{equation}
 \xi^V_{\alpha H}(\vxperp, v) = { \int  dv' /c\,
 \int_{N_{LL}}^\infty dN f(N) V(N, v'-v) \xi_{\alpha H}(\vxperp, v') \over
 \int_{N_{LL}}^\infty dN f(N) W(N) / \lambda_\alpha } ~,
\end{equation}
where $W(N)$ is the rest-frame equivalent width of an absorber of
column density $N$, and $N_{LL}=10^{17.2}\cm^{-2}$. The linear form for
the cross-correlation
$\xi_{\alpha H}$ in redshift space is readily obtained from the 
equations in \cite{HAMIL92}, replacing $\beta$ for
$(\beta_\alpha+\beta_H)/2$ and $\beta^2$ for $\beta_\alpha \beta_H$,
and the convolution in the above equation can be computed and used
in the model to fit the observed correlation function, by approximating
$\xi_F\simeq \xi_2$ and replacing $\xi_{\alpha H}$ by
$\xi^V_{\alpha H}$ in equation (\ref{eq:xi2def}).
Although we have not carried out this procedure for our mocks in this
paper, our analysis in this section and the results in Figure \ref{fig:terms}
indicate that if this improved model is directly fitted to the data,
the effects of HCDs ought to be corrected for to a high accuracy.

A similar approach was used in the Appendix B of \cite{2011MNRAS.415.2257M}.

%Analyzing this effect carefully is beyond the scope of this paper,
%but we believe that this is the principal reason for the discrepancy
%between our analytical estimate and the numerical calculation of the
%correction to the bias factors.

%% file: Metals.tex
\section{Effect of Metal Lines}
\label{sec:metals}

  Apart from HCDs, the other contaminating absorption that is found in
the spectral range where the \lya forest is observed is from absorption
lines of ionized heavy elements, or metals. When the transmission
fraction is measured at a certain pixel, part of the absorption may be
caused by any of the numerous intergalactic metal lines that are present
in the ultraviolet spectrum. For metal lines with wavelengths close to
that of the hydrogen \lya line, their absorption can overlap with
nearby, and therefore correlated, \lya absorption. This introduces new
components to the total transmission correlation, which will appear as
peaks in the three-dimensional correlation centered on the line of
sight at a velocity separation corresponding to the rest-frame
wavelength separation of the metal and \lya line. In the same way,
pairs of metal lines overlapping the \lya forest will also introduce
peaks in the correlation function at the wavelength separation of each
pair of metal lines.

  Although we shall not treat the impact of metals in detail in this
paper, we introduce here the general formalism for metal line
corrections to the overall transmission correlation that is analogous to
the one used for HCDs, which we believe will be useful for treating
their effect. Let the transmitted fraction field of each one of the
metal lines, $i$, that may appear in the spectral region of the \lya
forest, be $F_i = \bar F_i (1+\delta_i)$. In this section, the \lya
transmission is not separated into a \lya forest and HCD part, for
simplicity. The total transmission fraction field that is measured is
\begin{equation}
 F = F_{\alpha}\, \prod_i F_i ~ ;
\end{equation}
\begin{equation}
 \bar F \, (1+\delta_F) = \bar F_{\alpha}\, \prod_i \bar F_i ~
 (1+\delta_{\alpha}) \, (1+\delta_i) ~ .
\end{equation}

  We now define the following two-point cross-correlations:
\begin{equation}
 \xi_{\alpha i}({\bf x}) = \,
 \left< \delta_\alpha({\bf r}) \delta_i({\bf r} + {\bf x} ) \right> +
 \left< \delta_\alpha({\bf r}) \delta_i({\bf r} - {\bf x} ) \right> ~ ;
\end{equation}
\begin{equation}
 \xi_{ij}({\bf x}) = \,
 \left< \delta_i({\bf r}) \delta_j({\bf r} + {\bf x} ) \right> +
 \left< \delta_i({\bf r}) \delta_j({\bf r} - {\bf x} ) \right> ~ .
\end{equation}
As usual, brackets denote ensemble averages over all pixel pairs
separated by $\vx$. Let $\lambda_i$ be the central wavelength of each
metal line $i$, and let the vector ${\bf x}_i$ be directed along the
line of sight with its radial component equal to
$x_i = cH^{-1} (\lambda_i - \lambda_{\alpha})/\lambda_{\alpha}$. The
symmetrized cross-correlation $\xi_{\alpha i}$ then has two peaks, at
$\vx = \pm {\bf x}_i$, corresponding to the possibilities of having the
\lya line near $\vR$ and the metal line near $\vR + \vx$, or the \lya
line near $\vR - \vx$ and the metal line near $\vR$ [and two peaks at
$\pm(\vx_i-\vx_j)$ for $\xi_{ij}$]. In the linear regime, these
cross-correlations are simply the Fourier transforms of analogous
power spectra to those in equations (\ref{eq:powh}) and
(\ref{eq:powah}), which can be modeled in terms of bias factors and
redshift distortion factors for each metal line, $b_i$ and $\beta_i$,
defined as in equations (\ref{eq:biash}) and (\ref{eq:betah}), and
related to the bias factors of their host halos and their
mean effective optical depth in full analogy to equations
(\ref{eq:bhrel}) for the HCDs. Therefore, the two-point correlations
$\xi_{\alpha i}$ and $\xi_{ij}$ are fully specified by these factors
as long as the linear regime approximation is valid.

  We define also:
\begin{equation}
 C_m = \bar F / (\bar F_\alpha \prod_i \bar F_i ) =
 1 + \sum_i \xi_{\alpha i}(0) + \sum_{ij} \xi_{ij}(0) + ... ~ ,
\end{equation}
where we have omitted the three-point and higher-point functions. Note
that $C_m$ is nearly equal to unity, because the displacements $x_i$ are
usually large (the largest contributions to $C_m$ should arise from
lines with a wavelength close to that of Ly$\alpha$, e.g., from SiIII),
so the correlations at zero separation are very small.

  The observed absorption perturbation field $\delta_F$ is then
\begin{equation}
 C_m \delta_F = (1+ \delta_\alpha) \prod_i (1+\delta_i) - C_m ~ .
\end{equation}
Evaluating the correlation of $\delta_F$, and keeping only the most
important terms, which are the two-point functions and the three-point
term that contains only one metal contribution, we have
\begin{align}
\nonumber C_m^2 \xi_F ({\bf x}) =  & - (C_m-1)^2 + \xi_\alpha({\bf x}) +
 \sum_i \xi_{\alpha i}({\bf x}) + \sum_i \xi_{ii}({\bf x}) +
 \sum_{i<j} \xi_{ij}({\bf x}) + \\
 & + \sum_i \left< \delta_\alpha({\bf r}) \delta_\alpha({\bf r} \pm {\bf x}) 
          \delta_i({\bf r} ) \right> 
\end{align}
where the symbol $\pm$ implies here that there are two terms in the sum,
one with the $+$ sign and one with the $-$ sign.

  Ignoring the last three-point term, which is probably very small (and
may be computed numerically in simple models where metal lines are added
in mocks with a similar prescription as the one we have used for HCDs,
and can probably also be modeled as a product of the two correlations
$\xi_{\alpha i}$ and $\xi_\alpha$), this shows that the effect of every
metal line is fully specified by the parameters $b_i$ and $\beta_i$,
which can be related to the physical bias parameters of the host halos
through the mean effective optical depth of every line,
$\bar\tau_{ei}$. All these parameters should be directly measurable
from the data, by extracting the shape of the peaks of the overall
transmission correlation near the peak positions $\vx_i$. Obviously
these measurements will not be possible for weak metal lines as the
peaks they create become buried into the noise, but it should still
be possible to obtain combined measurements of the parameters for a
set of metal lines that are assumed to be hosted by the same halos
(and therefore have the same values of $b_i'$ and $\beta_i$ as in
equations \ref{eq:bhrel}). This leads to a proposed outline for a
program to be carried out to investigate all of the metal lines that
can be statistically detected: to measure their parameters
$\bar\tau_{ei}$, $b_i$ and $\beta_i$ and to use the formalism presented
here to correct for their impact on the overall transmission correlation
function. 

%NOTE: comment on Matt's stacks, and bumps in the 1D correlation measured
%both by Pat in 2006 and by the FPG with BOSS data.

%% file: Conclusions.tex
\section{Conclusions}

  The measurement of the bias and redshift distortion parameters of the
\lya forest correlation function may reveal essential characteristics of
the physical evolution of the intergalactic medium. However, their
values are affected by the presence of the absorption profiles from high
column density systems (HCDs) and metal lines in the observed spectra.
It is therefore necessary to study how these systems affect the
correlation function to try to correct the measured parameters for their
effect.

  We have presented a numerical method to simulate the effect of HCDs
(HCD) on the measured correlation function of the \lya absorption, in
which the Voigt profiles of absorbers are inserted in mock spectra of
the \lya forest in positions that are correlated with the \lya optical
depth. We have evaluated the increase of the noise in the correlation
measurement and the systematic change that is introduced in the
recovered bias parameters owing to the correlation
of HCDs with the \lya forest. Both effects are very substantial: the
HCDs contribution to the noise is close to that of the \lya forest
itself, and the bias parameters are altered by $\sim 10\%$ in the
models we have used. Even though most of the increase in the
noise is caused by damped \lya systems that can individually be
identified in the spectra, and therefore removed to reduce the noise,
this should be considered as a dangerous operation to do because the
identified systems that can be detected may have different bias factors,
and therefore cause different systematic effects, than the set of all
HCDs. If the removal of detected HCDs is attempted (either by masking
regions of the spectrum where HCDs are present or fitting their profiles
as part of the continuum), one should examine the difference in
the results of the \lya correlation measurements when no HCDs are
removed, and test the selection of HCDs and their effects through
simulations using mock spectra in which HCDs are inserted with the
observed correlation with the \lya forest.

  An analytical formalism has also been
developed to more generally predict the changes induced by HCDs on the
inferred linear bias parameters. We find that the most important terms
biasing the correlation function can be computed from the two-point
cross-correlations of the \lya and HCD absorption perturbation fields.
Assuming the validity of linear theory and that HCDs are associated
with their host halos in a way that is independent of the large-scale
environment, we have inferred a set of simple equations relating
the corrections on the bias factors to the effective optical depth and
the host halos bias factor of the HCDs. The density bias factor is
increased in absolute value by the product of the mean effective
optical depth and the bias factor of the host halos of the HCDs
(eq.\ \ref{eq:db1}), and the redshift distortion parameter is altered
also in proportion to $\bar\tau_{eH}$ and the difference
$\beta_\alpha - \beta_H$ (eq.\ \ref{eq:db3})

  Even though the results that can be derived analytically go in the
direction of the correction to the fitted density bias factor that we
find numerically for the specific model of the mocks we have analyzed,
they do not quantitatively agree. We have identified two reasons for
this discrepancy: one is the neglect of the three-point and four-point
terms in the analytical approach, in particular the term $\xi_{3\alpha}$
in equations (\ref{eq:lyadla}) and (\ref{eq:termdef}). The other one is
the deviation from the linear theory form of the cross-correlation of
the \lya forest and HCD absorption due to the extended nature of the
damped absorption profiles. The latter effect, in particular, should
mostly explain the change in the total flux redshift distortion
parameter, $\beta_F$, in our mocks (see Table 1), despite the fact that
by construction, the HCDs are inserted with a distribution obeying
$\beta_H = \beta_\alpha$. The damped wing profiles are acting analogously
to ``fingers of God'' in galaxy redshift surveys to distort the contours
of the correlation function and reduce the fitted value of $\beta_F$.

  The results of this paper lead us to believe that it is possible to
accurately correct for the contamination introduced by HCDs in the
total transmission correlation function $\xi_F$, to infer the true
underlying \lya correlation function $\xi_\alpha$, and therefore to
constrain our models for the evolution of the intergalactic medium.
In principle, by simply convolving the linear theory form of the
cross-correlation function $\xi_{\alpha H}$ with a Voigt profile in
the radial direction that results from the mean column density in HCDs
associated with a \lya forest transmission perturbation
$\delta_\alpha$, one should have a much more accurate model to be
fitted to the observations. The effect of the next largest term,
$\xi_{3\alpha}$, can be numerically taken into
account once the parameters for the HCD model (mainly $b_H$, $\beta_H$
and $\bar\tau_{eH}$) have been calibrated through the observational
determination of the \lya forest - HCD cross-correlation.

%In their Appendix B, \cite{2011MNRAS.415.2257M} presented a toy model to
%study the effect of HCDs on the 3D power spectrum of the \lya forest.
%They found that the systems have a contribution of $\sim 60 \%$ on large
%scales, significantly larger than the results presented here. The 
%discrepancy can be explained by the effect of redshift space distortions 
%(not included in their study), since they add more signal in the \lya power 
%than in that of HCDs.
%In the same study, the authors claimed that there is little point of masking 
%the larger systems ($N_{HI} > 10^{20}\cm^{-2}$) since they contribute a small
%component of the total power. Since in their model the contribution is 
%proportional to the fraction of flux absorbed by the systems, this statement
%would imply that most of the absorption comes from lower column densities, 
%in disagreement with our Figure \ref{fig:FH}.

  Although we have not attempted an evaluation of the impact of metal
lines on the correlation function in this paper, we have proposed a
basic formalism to understand and correct for their effect that is
similar to the one for HCDs. For the metal lines that contaminate the
spectral region of the \lya forest, the metal-\lya cross-correlations
can also be measured from the data and its effect included when the
observations of the overall flux correlation $\xi_F$ are fitted.
We are therefore optimistic on the prospects for obtaining accurate
measurements of the \lya correlation as a probe to the large-scale
primordial perturbations and the physical evolution of the intergalactic
medium from the BOSS and other future spectroscopic surveys of the \lya
forest.

%% file: acknowledgements.tex
\section*{Acknowledgements}

  We would like to thank Patrick McDonald, Matt McQuinn, Uros Seljak, 
Patrick Petitjean, and Anze Slosar for many illuminating discussions.
We also thank CITA for their hospitality and the use of their computer
resources by A. Font-Ribera to accomplish part of this work. This
work was supported in part by Spanish grants AYA2009-09745 and
PR2011-0431.

%% file: Peaks_bias.tex
\section{Appendix: Clustering of the HCD systems}
\label{app:bias}

In this Appendix we show that the method described in Section \ref{sec:mockdla}
distributes the HCDs following a correlation function $\xi_h(\vR)$ in redshift
space that is, on large scales, proportional to the flux correlation
function $\xi_F(\vR)$.
This implies that the redshift space distortion parameters of the two
correlations are equal, $\beta_h=\beta_\alpha$. Furthermore,
the ratio of these two correlations yields the relation between the bias
factors of the distribution of HCDs and of the transmitted flux fraction
field, $b_h$ and $b_F$,
\begin{equation}
 \frac{\xi_h(\vR)}{\xi_F(\vR)} = \left(\frac{b_h}{b_F} \right)^2 ~.
\end{equation}
%Because this relation is computed directly in redshift space, the correlation 
%function of both fields have the same shape on large scales, i.e. the same
%redshift space distortion parameter $\beta_h = \beta_F$.

  To simplify our notation in this Appendix, we use $F$, $\bar F$ and
$\xi_F$ to refer to the \lya forest variables that in the main text
are referred to as $F_\alpha$, $\bar F_\alpha$ and $\xi_\alpha$.

 The method to generate the mock \lya absorption field, described in detail in 
\cite{2012JCAP...01..001F} and summarized in Section \ref{sec:mockdla},
generates first a random Gaussian field $\delta_g$ with a correlation function
$\xi_g(\vR)$, such that the final flux field
$F(\delta_g)$ has the desired correlation $\xi_F(\vR)$, as well as the
desired probability distribution function $p_F(F)$.
We first prove that this auxiliary Gaussian field has the 
same correlation function as the flux field, with a different bias 
parameter $b_g$. Then we show that the peaks of the Gaussian field also
have the same correlation function,
with a relative bias set by the peak threshold.

\subsection{Biases of the Gaussian field}

  The relation between the correlation of any function $F(\delta_g)$
and $\xi_g$ can be computed as follows:
\begin{align}
\nonumber \bar{F}^2 [1+ \xi_F(\vR_{12})]
 & = \left< \, F(\vx_1) \, F(\vx_2) \,\right> 
       = \int_0^1 dF_1  \int_0^1 dF_2 \, p_F(F_1,F_2) \, F_1 \, F_2 \\
\nonumber   & = \int_{-\infty}^\infty d\delta_{g1}  \int_{-\infty}^\infty 
                       d\delta_{g2} ~p_g(\delta_{g1},\delta_{g2})
			\, F(\delta_{g1}) \, F(\delta_{g2}) \\
            & = \int_{-\infty}^\infty d\delta_{g1}  
                  \int_{-\infty}^\infty d\delta_{g2} \,
                  \frac{ \exp \left[-\dfrac{\delta_{g1}^2 + \delta_{g2}^2
                                -2\delta_{g1} \delta_{g2} \xi_g(r_{12})}
       {2\left[ (1-\xi^2_g(\vR_{12}) \right]} \right]}
                       {2 \pi \sqrt{1-\xi^2_g(\vR_{12})}} 
                  ~F(\delta_{g1}) ~F(\delta_{g2}) ~,
\label{eq:xiff}
\end{align}
In the method to generate the \lya mocks, this expression is inverted to
find $\xi_g$ as a function of the desired $\xi_F$. In this Appendix
we use it to show that, on large scales, the two correlation functions
are proportional to each other.

The Gaussian variables $\delta_{g1}=\delta_g({\bf x}_1)$ and 
$\delta_{g2}=\delta_g({\bf x}_2)$ are normalized to unit dispersion
and their correlation is $\xi_g(\vR_{12})$. 
We define the new normal variables $y_1$, $y_2$ as linear combinations
that are independent:
\begin{equation}
 \label{eq:y}
  \delta_{g1} = y_1 ~;   \qquad   
  \delta_{g2} = \xi_g \, y_1 + \sqrt{1-\xi_g^2} \, y_2 ~.
\end{equation}
In the linear regime, we can assume $\xi_g \ll 1$ and use a
first-order expansion of the function $F(\delta_{g2})$,
\begin{equation}
 F( \delta_{g2}) \approx F(y_2) + \frac{dF}{d\delta_g} \, y_1\, \xi_g
      = F(y_2) \left( 1 - \frac{d\tau}{d\delta_{g2}} \, y_1\, \xi_g \right) ~ , 
\end{equation}
where $\tau = -\,{\rm ln}(F)$.

The flux correlation in equation (\ref{eq:xiff}) now becomes,
\begin{equation}
 \bar{F}^2 \left[1 + \xi_F(\vR_{12}) \right] \approx 
   \int_{-\infty}^\infty dy_1 \, p_g(y_1) \, F(y_1)  
      \int_{-\infty}^\infty dy_2 \, p_g(y_2)
            \, F(y_2) \left(1-\frac{d\tau}{dy_2} \, y_1 \, \xi_g \right) ~,
%  & = \bar{F}^2 \left[ 1 + \left( \frac{b_F}{b_g} \right)^2 \xi_g \right]  ~.
\end{equation}
where $p_g(y)$ is the one-dimensional, normal Gaussian distribution.
Requiring now that $\xi_F(\vR_{12}) = (b_F/b_g)^2 \xi_g(\vR_{12})$,
we find that the ratio of bias parameters is fully determined by
the transformation $F(\delta_g)$:
\begin{equation}
 \left( \frac{b_F}{b_g} \right)^2 = \frac{1}{\bar{F}^2} 
    \int_{-\infty}^\infty dy_1 \, p_g(y_1) \, F(y_1) \, y_1 \,  
    \int_{-\infty}^\infty dy_2 \, p_g(y_2) \, F(y_2) \, \frac{d\tau}{dy_2} ~.
\end{equation}

The mocks used in this paper are computed using a lognormal
transformation for the optical depth $\tau$,
\begin{equation}
 F(\delta_g) = \exp \left[ -\tau(\delta_g) \right] 
    = \exp \left[ -a e^{\gamma \delta_g} \right] ~,
\end{equation}
with $a=0.077$ and $\gamma=2.16$. Using this transformation, the ratio of
the bias factors is
\begin{equation}
  \left( \frac{b_F}{b_g} \right)^2 = 0.0925 ~.
\label{eq:biasr}
\end{equation}

\subsection{Biases of the peaks}

In section \ref{sec:mockdla} we describe a method to distribute HCD systems 
in regions of the \lya spectra where the optical depth is above a
threshold $\tau_c$ or, equivalently, above a threshold 
$\delta_{gc}$ in the Gaussian variable used to generate the optical depth. 
We refer to these regions as {\it peaks}.
The threshold sets the fraction $\nu$ of pixels that are candidates to
host a HCD:
\begin{equation}
  \nu = \int_{\delta_{gc}}^\infty d\delta_{g} \, p_g(\delta_g) ~.
\end{equation}

The correlation function of these peaks, $\xi_h$, is related to the probability
of having a peak both at $\delta_{g1}=\delta_g(\vx_1)$ and at 
$\delta_{g2} =\delta_g(\vx_2)$:
\begin{align}
  p(\delta_{g1}>\delta_{gc} , \, \delta_{g2}>\delta_{gc}) 
\nonumber     & = \nu^2 ~ \left[ 1 + \xi_h(\vR_{12}) \right] \\
\nonumber     & = \int_{\delta_{gc}}^\infty d\delta_{g1}  
                  \int_{\delta_{gc}}^\infty d\delta_{g2} 
                  \, p_g(\delta_{g1} , \delta_{g2}) \\
              & = \int_{\delta_{gc}}^\infty d\delta_{g1}  
                  \int_{\delta_{gc}}^\infty d\delta_{g2} 
                \, \frac{ \exp \left[ -\dfrac{\delta_{g1}^2 
                  + \delta_{g2}^2-2\delta_{g1}\delta_{g2}\xi_g(\vR_{12})}
                                  {2[1-\xi^2_g(\vR_{12})]} \right]} 
                       {2 \pi \sqrt{1-\xi^2_g(\vR_{12})}}  ~.
\end{align}

We now express $\delta_{g1}$ and $\delta_{g2}$ as a function of the
independent normal variables $y_1$ and $y_2$ defined in equation \ref{eq:y}.
Defining $\delta_{gc}^\prime (y_1,\xi_g) = (\delta_{gc} - y_1 \xi_g)/
 (1-\xi_g^2)^{1/2}$, we obtain
\begin{align}
 \nonumber \nu^2 ~(1 + \xi_h) & = \int_{\delta_{gc}}^\infty dy_1 \, p_g(y_1) 
                        \int_{\delta^\prime_{gc}(y_1,\xi_g)}^\infty
                               dy_2 \, p_g(y_2) \\
\nonumber     & = \int_{\delta_{gc}}^\infty dy_1 \, p_g(y_1) \times
                    \left[ \int_{\delta_{gc}^\prime(y_1,\xi_g)}^{\delta_{gc}}
			 dy_2 \, p_g(y_2)
                    +  \int_{\delta_{gc}}^\infty dy_2 \,p(y_2) \right] \\
\nonumber     & \approx \int_{\delta_{gc}}^\infty dy_1 \, p_g(y_1) 
            \left\{\left[ \delta_{gc} - (\delta_{gc} - y_1 \, \xi_g) \right]
                        \, p_g(\delta_{gc}) + \nu\, \right\} \\
       & = \nu^2 \left[ 1 + \left( \frac{b_h}{b_g} \right)^2 \xi_g \right] ~,
\end{align}
where $b_g$ and $b_h$ are the bias parameters of the Gaussian field and 
the peaks, respectively. The bias ratio is then a function of $\nu$ only:
\begin{equation}
 \left( \frac{b_h}{b_g} \right)^2 = \frac{p_g(\delta_{gc})}{\nu^2} 
                 \int_{\delta_{gc}}^\infty dy_1 \, p_g(y_1) \, y_1 ~.
\end{equation}
For the values used in the models in this paper, we find $b_h/b_g=2.66$
for $\nu=0.01$, and $b_h/b_g=3.18$ for $\nu=0.002$. Using
equation (\ref{eq:biasr}) and the value of the \lya forest bias in our
mocks, $b_F=-0.1375$, we find the bias of the HCD systems to be
$b_h=1.21$ for $\nu=0.01$, and $b_h=1.43$ for $\nu=0.002$.

  Finally, because the correlation of the peaks has been shown to be
proportional to the correlation of the Gaussian field, their redshift
distortion parameter must be the same.